\documentclass[fleqn,usenatbib]{mnras}

\usepackage{newtxtext,newtxmath}

\usepackage[T1]{fontenc}

\DeclareRobustCommand{\VAN}[3]{#2}
\let\VANthebibliography\thebibliography
\def\thebibliography{\DeclareRobustCommand{\VAN}[3]{##3}\VANthebibliography}

\usepackage{graphicx}	
\usepackage{amsmath}	
\usepackage{hyperref}
\usepackage{bm}

\DeclareMathOperator{\Tr}{Tr}

\def\para{\parallel}

\def\be{\bm{e}}
\def\bv{\bm{v}}
\def\bV{\bm{V}}
\def\bJ{\bm{J}}
\def\bk{\bm{k}}
\def\br{\bm{r}}
\def\bx{\bm{x}}
\def\bz{\bm{z}}

\def\bR{\bm{R}}
\def\bW{\bm{W}}
\def\bp{\bm{p}}
\def\bq{\bm{q}}
\def\bM{\bm{M}}
\def\bH{\bm{H}}
\def\bM{\bm{M}}
\def\bH{\bm{H}}
\def\bSigma{\bm{\Sigma}}
\def\bd{\bm{d}}
\def\bmm{\bm{m}}

\def\bs{\bm{s}}
\def\btheta{\bm{\theta}}
\def\bB{\bm{B}}
\def\bA{\bm{A}}
\def\bG{\bm{G}}

\def\bC{\bm{C}}
\def\bL{\bm{L}}

\title[Differentiable Bayesian Forward Modeling for 21\,cm Cosmology]{A Differentiable, End-to-End Forward Model for 21\,cm Cosmology: \\
Estimating the Foreground, Instrument, and Signal Joint Posterior}

\author[N. Kern]{
Nicholas Kern$^{1,2}$\thanks{NASA Hubble Fellow; E-mail: nkern@umich.edu}
\\
$^{1}$Department of Physics, University of Michigan, Ann Arbor , MI\\
$^{2}$MIT Kavli Institute, Massachusetts Institute of Technology, Cambridge, MA
}

\date{Accepted XXX. Received YYY; in original form ZZZ}

\pubyear{\the\year{}}

\begin{document}
\label{firstpage}
\pagerange{\pageref{firstpage}--\pageref{lastpage}}
\maketitle

\begin{abstract}
We present a differentiable, end-to-end Bayesian forward modeling framework for line intensity mapping cosmology experiments, with a specific focus on low-frequency radio telescopes targeting the redshifted 21\,cm line from neutral hydrogen as a cosmological probe.
Our framework is capable of posterior density estimation of the cosmological signal jointly with foreground and telescope parameters at the field level.
Our key aim is to be able to optimize the model's high-dimensional, non-linear, and ill-conditioned parameter space, while also sampling from it to perform robust uncertainty quantification within a Bayesian framework.
We show how a differentiable programming paradigm, accelerated by recent advances in machine learning software and hardware, can make this computationally-demanding, end-to-end Bayesian approach feasible.
We demonstrate a proof-of-concept on a signal recovery problem for the Hydrogen Epoch of Reionization Array experiment, highlighting the framework's ability to build confidence in early 21\,cm signal detections even in the presence of poorly understood foregrounds and instrumental systematics.
We use a Hessian-preconditioned Hamiltonian Monte Carlo algorithm to efficiently sample our parameter space with a dimensionality approaching $N\sim10^5$, which enables joint, end-to-end nuisance parameter marginalization over foreground and instrument terms.
Lastly, we introduce a new spherical harmonic formalism that is a complete and orthogonal basis on the cut sky relevant to drift-scan radio surveys, which we call the spherical stripe harmonic formalism, and it's associated three-dimensional basis, the spherical stripe Fourier-Bessel formalism.
\end{abstract}




\section{Introduction}

One of the frontiers of modern astrophysics and cosmology is the study of the high-redshift universe, particularly the epochs between the emission of the Cosmic Microwave Background (CMB) at recombination ($z\sim1100$) and the onset of dark energy driven expansion ($z\sim0.5)$.
While the early universe and late universe have been mapped in exquisite detail, constraining the age, structure, and expansion of the universe \citep{Planck2020, DES2022, DESI2024}, the intervening epochs have been relatively underexplored.
In particular, our understanding of the birth of the first stars, galaxies, and black holes, known as Cosmic Dawn, is only weakly constrained.
Integrated CMB measurements combined with quasar absorption and galaxy observations tell us that the Epoch of Reionization (EoR), which marks the ionization of neutral hydrogen in the intergalactic medium (IGM) from early stellar populations, is ending around a redshift $z\sim6$ \citep{Robertson2015, Mason2018, Davies2024}.
Furthermore, recent observations from the James Webb Space Telescope (JWST) have revealed some of the brightest galaxies emerging from Cosmic Dawn; however, these observations have also complicated our understanding of the growth of galaxies and the total ionizing photon budget of the first stellar populations \citep{Robertson2023, Yung2024, Munoz2024}.
Thus, alternative probes of the high-redshift universe, in particular ones that can reach deep into the early stages of Cosmic Dawn, are vital for constructing a comprehensive understanding of high-redshift astrophysics.

Mapping cosmologically-redshifted hydrogen via its 21\,cm emission, known as 21\,cm cosmology, has long been known as a potentially transformative probe of cosmology and astrophysics.
It is a direct probe of the IGM during Cosmic Dawn and the Dark Ages, and is sensitive to inflationary physics \citep{Scott1990, Loeb2004, Mao2008} and the wide landscape of astrophysical models governing the formation of the first stellar populations and their feedback on the IGM \citep[e.g.][]{Madau1997, Furlanetto2006c, Morales2010, Pritchard2012, Mesinger2012, Fialkov2014b, Liu2020}.
In the post-reionization era at $z < 5.5$, the 21\,cm line is a tracer of the density field on large scales capable of constraining cosmological structure growth and deviations from $\Lambda$CDM cosmlogy \citep{Shaw2014, Bull2015, Obuljen2018}.
However, across the redshift spectrum, 21\,cm cosmology radio surveys are hindered by exceedingly bright astrophysical foregrounds that dwarf the cosmological signal by upwards of a factor of $10^{10}$ in the power spectrum \citep{Liu2020}.
This results in an exceedingly difficult signal separation problem, which has thus far made a robust direct detection of the 21\,cm signal elusive.

Nonetheless, a wide range of experimental efforts have made tremendous progress over the past decade in setting increasingly stringent limits on the cosmological 21\,cm signal.
This includes probes of the Cosmic Dawn 21\,cm power spectrum \citep{Paciga2013, Trott2020, HERA2022a, HERA2023, Munshi2024, Mertens2025, Nunhokee2025}, the Cosmic Dawn 21\,cm monopole \citep{Bernardi2016, Bowman2018, Singh2018b}, and the post-reionization neutral hydrogen signal \citep{Chang2010, Masui2013, Paul2023, CHIME2024}.
Recent upper limits on the Cosmic Dawn 21\,cm power spectrum from the Hydrogen Epoch of Reionization Array \citep[HERA;][]{DeBoer2017, Berkhout2024} have placed the most stringent constraints on the heating of the high-redshift IGM at $z>8$ and the efficiency of the first X-ray emitters \citep{HERA2022a, HERA2022b, HERA2023}.

Going forward, how we transition from setting upper-limits on the 21\,cm power spectrum to making a direct detection is more complex.
A suite of tools have been developed for residual systematic testing \citep{HERA2022a, Wilensky2023} and for simulated pipeline validation on data mocks \citep{Barry2019a, Mertens2020, Hothi2020, Tan2021, Aguirre2022, Line2025}, which will help build confidence in early detections.
However, we currently lack a framework for inverting the effects of systematics in an end-to-end fashion, and furthermore lack the ability to propagate the uncertainty on these terms to our final inferences in a statistically robust manner.
Recently, the importance of end-to-end modeling for line intensity mapping (LIM) surveys has been appreciated, with particular emphasis placed on more realistic systematic modeling \citep{Aguirre2022, Fronenberg2024, Cheng2024, Kittiwisit2025, Ohara2025}.
Nevertheless, an end-to-end model that is capable of mitigating the combined effects of foregrounds and systematics in raw 21\,cm datasets to the required precision 
 is currently lacking.

Another way of phrasing the problem, from a Bayesian perspective, is that we currently lack a robust way to estimate the joint posterior distribution between the 21\,cm signal, astrophysical foregrounds, and instrumental systematics.
In theory, a robust power spectrum detection would entail marginalizing over the foreground and systematic nuisance parameters to yield a marginal posterior distribution that accounts for uncertainties due to thermal noise fluctuations \emph{in addition to} the intrinsic degeneracies between the 21\,cm signal and various systematics.
End-to-end pipelines are key to this process, as they allow us to propagate subtle effects through our complex data model.
Indeed, end-to-end approaches are increasingly being deployed for astrophysical and cosmological analyses where systematics are a major limiting factor \citep[e.g.][]{BeyondPlanck2023, Alsing2023, Popovic2023}.

Bayesian approaches to signal separation problems in cosmology found early traction in CMB data analysis \citep[e.g.][]{Jewell2004, Wandelt2004, Eriksen2004, Eriksen2008}.
Since then, the advent of automatic differentiation (AD) and backpropagation methods for computing gradients of non-linear, black-box models \citep{Baydin2015} has led to the wider adoption of end-to-end Bayesian forward modeling in cosmological data analysis \citep[e.g.][]{Jasche2013, Horowitz2021, Bohm2021, Gu2022, Hahn2023, Li2024, Nguyen2024}.
This adoption has been fueled both by user-friendly AD-enabled software frameworks \citep[e.g.][]{Campagne2023, Li2024}, but also by the advent of large-memory graphics processor unit (GPU) computing that excels in accelerating the kind of matrix operations central to scientific computing.

Given the difficult signal separation problem facing 21\,cm cosmology, a fresh wave of attention has been given to Bayesian methods in recent years \citep[e.g.][]{Zhang2016, Mertens2018, Sims2019, Rapetti2020, Anstey2021, Byrne2021, CHIME2022, Burba2023, Kennedy2023, Anstey2023, Scheutwinkel2023, Burba2024, Murphy2024, Pagano2024, Glasscock2024}.
In particular, approaches for Bayesian modeling of 21\,cm datasets have been proposed and shown to acheive compelling results for estimating foreground and 21\,cm signal parameters jointly on simulated datasets.
The approach in \citet{Sims2019, Burba2023} explores this question in the context of sampling the joint 21\,cm and foreground power spectrum, but has thus far been confined to conditioning on the instrumental response.
\citet{CHIME2022, Kennedy2023, Burba2024} take a Gibbs sampling approach, reconstructing the power spectrum jointly with the data under various contamination scenarios, but operate solely at the visibility level and thus are not fully end-to-end models.
\citet{Glasscock2024} expands upon this, proposing an end-to-end Gibbs sampler for reconstructing low-angular resolution diffuse foreground maps.
However, to date, an end-to-end forward model capable of jointly sampling the 21\,cm signal, foregrounds, and the instrumental response is lacking.

In this work, we present the first end-to-end Bayesian forward model employing modern automatic differentiation techniques for 21\,cm cosmological analysis, called \texttt{BayesLIM}.\footnote{\url{https://github.com/BayesLIM/BayesLIM}} 
It is capable of estimating the joint posterior between the foreground sky, the instrumental response, and the three-dimensional 21\,cm sky signal at the field level.
It is a highly flexible and modular code designed to tackle a wide range of problems found in practical LIM scientific analysis.
The framework parameterizes sky signals as 3D fields and numerically computes the telescope measurement process, adding in instrumental corruptions along the way.
Expressing our forward model in an automatically differentiable programming language \citep{pytorch} enables fast and efficient parameter gradient calculations.
This in turn allows us to leverage optimization and Markov Chain Monte Carlo (MCMC) samplers that are particularly efficient for high-dimensional problems, such as quasi-Newton optimizers and Hamiltonian Monte Carlo (HMC) samplers.
Furthermore, the easy GPU-portability afforded by modern differentiable programming languages helps to accelerate the computationally intensive end-to-end forward model approach.

This framework is applicable to both interferometric and total-power intensity mapping surveys.
In addition, while it is currently tuned for 21\,cm intensity mapping, it is in principle a general framework capable of modeling and synthesizing together multiple intensity mapping probes.
The challenge of such an approach mainly lies in accelerating the forward model such that it can be reasonably evaluated on the order of thousands of times or more, and the large memory footprint created by the computational graph.
The former is alleviated by GPU acceleration, while the latter can be addressed by making judicious parameterization choices, in addition to standard techniques like gradient accumulation, data parallel training, and gradient checkpointing.
Indeed, the recent availability of high-performance, large-memory GPU compute is key to enabling the approach described in this work.

To demonstrate our framework, we apply it to a mock observation for the Hydrogen Epoch of Reionization Array (HERA) experiment.
For simplicity in this proof-of-concept work we only consider the joint modeling of: 1. the wide-field foreground sky, 2. the (antenna-independent) horizon-to-horizon antenna primary beam response, and 3) the 21\,cm sky signal.
In total, our model contains roughly 80,000 active parameters spanning those three components.
Note that the ultimate goal is to not only produce maximum a posteriori (MAP) inference of the 21\,cm signal, but also to explore the inherent degeneracies between the foregrounds, instrument, and 21\,cm signal parameters, thereby estimating the joint posterior of the model at the field level.
Future work will explore how to include other instrumental parameters such as antenna gain calibration and mutual coupling \citep{Kern2020a, Josaitis2022, Rath2024, Ohara2025}.

In this paper we first discuss the 21\,cm cosmology inverse problem and the general forward modeling framework.
Next we describe in detail the choice of parameterization for our three model components and the mock observations used in this work.
Finally, we show the results of our forward model optimization and posterior sampling, demonstrating the first marginalized posterior distribution on the 21\,cm power spectrum from an end-to-end forward model across foreground and instrumental parameters.
Lastly, we derive a new spherical harmonic basis that is band-limited complete and orthogonal on the \emph{spherical stripe}, which is relevant for drift-scan 21\,cm surveys like HERA.
We call this the spherical stripe harmonics (SSH), and also discuss its associated 3D generalization, the spherical stripe Fourier Bessel (SSFB) formalism.

\section{Data Modeling Formalism}
\label{sec:formalism}
Here we describe our data modeling formalism for radio interferometric observations.
This includes a description of the forward model of the radio visibilities, which encodes the mapping of the model parameters to the observable data. We also discuss the data likelihood and model posterior distribution.


\subsection{The Radio Interferometric Measurement Equation}
\label{sec:rime}

The radio interferometric measurement equation (RIME) describes the fundamental measurable of a radio interferometer, known as the complex-valued \emph{visibility}, and relates it the response of the instrument and the radiation incident on it from the sky
\citep{Hamaker1996, Sault1996, Carozzi2009, Smirnov2011, Wilson2013}.
In brief, the RIME describes a series of operations that modulate celestial radiation and its polarization state as it travels to a radio antenna and is then converted into the visibility by correlating two antenna voltage streams.

Often the RIME is written in the flat-sky, small field-of-view (FoV) limit, in which case it can be shown that the radio visibilities are simply the two-dimensional Fourier trasnform of the sky brightness distribution weighted by the antenna primary beam response, which is also known as the van Cittert-Zernike theorem \citep{Wilson2013}.
However, in general, the RIME is the surface integral of the sky brightness distribution weighted by the antenna primary beam and the fringe response of a given baseline vector.
In this general form, the visibility for a baseline vector formed between two antennas $p$ and $q$ is written as
\begin{align}
\label{eq:rime_integral}
V_{pq}(\nu) = \int d^2\hat{\bm{s}}\ e^{-2\pi i \bm{b}_{pq}\cdot\hat{\bm{s}}\nu/c}\ A_{pq}(\hat{\bm{s}},\nu)\ B(\hat{\bm{s}},\nu),
\end{align}
where $\bm{b}_{pq} = \bm{r}_p - \bm{r}_q$ is the baseline vector, $\hat{\bm{s}}$ is the unit pointing vector of the surface integral decomposed in spherical coordinates into a polar unit vector $\hat{\theta}$ and an azimuthal unit vector $\hat{\phi}$, $A_{pq} = A_p A_q^\ast$ is the primary beam \emph{total power} response, assumed to be the same for all antennas, and $B$ is the unpolarized sky brightness distribution in units of specific intensity (Jansky/steradian).
For a drift-scan telescope, which points at a fixed location in topocentric coordinates as the Earth rotates, we can compute a unique visibility for each local sidereal time of our observations.
Thus the visibilities fundamentally have a baseline, frequency, and time dependence.

Note that \autoref{eq:rime_integral} is also defined for a single antenna feed polarization.
Typically a radio receiver will measure two orthogonal feed polarizations to reconstruct the full Stokes I distribution on the sky; however, in this proof-of-concept study we will restrict ourselves to a single feed polarization, which is generally a good approximation of the Stokes I power within the main field of view anyways \citep{Kohn2016}.

We can also incorporate the response of the telescope analog system (e.g. amplification) and electronics (e.g. analog-to-digital conversion) through what is called \emph{direction-independent} RIME terms, also known as the gain terms \citep{Smirnov2011}.
While this is an important component of a practical 21\,cm data analysis, we omit it here for brevity and only consider the antenna primary beam response as the instrumental component of our data model. Future work will explore joint modeling of gain and beam terms.
Note that we can easily incorporate polarized sky sources, multiple feed polarizations, and instrumental gain terms into a single RIME equation via its matrix-based Jones formalism \citep{Smirnov2011}.
However, given the limited scope of this proof-of-concept, we defer elaborating on this approach for future work.

If we discretize the integral into a sum over $N_{\rm pix}$ angular pixels, each with a solid angle $\delta\Omega$, we can write the numerical RIME as
\begin{align}
\label{eq:rime_discrete}
V_{\alpha\nu} = \delta\Omega \sum_j^{N_{\rm pix}}K_{j\alpha\nu}A_{j\alpha\nu}B_{j\nu},
\end{align}
where $K_{j\alpha\nu}=\exp[-2\pi i \bm{b}_\alpha\hat{\bm{s}}_j\nu/c]$, and $\alpha$ indexes each unique baseline ($pq$) in the array.
If we collect the sky brightness pixels into a vector and put the fringe and beam terms into a design matrix $\bm{A}\in\mathbb{C}^{N_{\rm baselines}\times N_{\rm pix}}$, then we can express the (noiseless) RIME as the linear model
\begin{align}
\label{eq:rime_matrix}
\bm{y} = \bm{A}\bm{x},
\end{align}
where $\bm{x}$ is a column vector of the pixelized sky, $\bm{y}$ is a vector of the measured visibilities for all baselines in the array, and the design matrix $\bm{A}$ is not to be confused with the primary beam response $A_{pq}$.
Here we've further assumed a celestial coordinates observer frame of reference, meaning that we have a unique $\bm{A}$ matrix for each observing time of the telescope.
Note that although \autoref{eq:rime_matrix} takes the form of a linear model, if we want to solve for different components within our forward model simultanouesly, for example the sky and the beam response, then we are left with a non-linear optimization problem.

A number of computer codes have been developed to efficiently evaluate the RIME for 21\,cm cosmology applications \citep[e.g.][]{Sullivan2012, Lanman2019c, Lanman2019b}, including GPU-accelerated codes \citep{Line2022, Kittiwisit2025, Ohara2025}.
The discretized surface integral approach in \autoref{eq:rime_matrix} is an exact model of point source emission; however, for extended emission like that from the galactic plane the discretization incurs an error.
One can make this error arbitrarily small by sampling at finer spatial resolutions.
The angular resolution of an inteferometric baseline with length $b$, observing at a wavelength $\lambda$, will have a spatial resolution of $\theta = \lambda / b$ radians.
Thus, we should discretize the sky at least as small as $\theta/2$ according to the Nyquist sampling theorem.
For this work, we use a central observing frequency of 125 MHz and a longest baseline of 60 meters, yielding an angular resolution of 2.3 degrees.
We discretize the sky using an equal-area, rectangular grid in declination and right ascension, with a pixel resolution of 0.5 degrees.

After simulating the model visibilities via \autoref{eq:rime_matrix}, we are free to apply any further operations to the data to aid in its comparison to the raw data.
It is common, for example, to filter the data across the frequency axis \citep[e.g.][]{Parsons2008, Mertens2020, Ewall-Wice2020, Kern2021} or across the time axis \citep[e.g.][]{Parsons2016, Kolopanis2019, Kern2020a, Garsden2024} to reduce the foreground contamination, or to perform baseline averaging to compress the data \citep{CHIMECollaboration2022, HERA2022a}. 
In this work, we will employ a high-pass delay filter on both the model visibilities and the noisy, raw visibilities to aid in comparing the two in our likelihood, which is applied to the simulated visibilities as
\begin{align}
\label{eq:rime_filter}
\bm{m} = \bm{F}\bm{y}^\prime
\end{align}
where $\bm{y}^\prime\in\mathbb{C}^{N_{\rm frequency}\times N_{\rm baselines}}$ is the stacked visibilities for all observing frequencies, $\bm{F}\in\mathbb{R}^{N_{\rm frequency}\times N_{\rm frequency}}$ is our high-pass filter, and $\bm{m}$ is our final model visibilities.
In the context of optimization, this filter helps to upweight the modes relevant to an EoR 21\,cm power spectrum detection, namely $k\gtrsim0.1\ {\rm Mpc}^{-1}$, relative to the otherwise dominating $k\sim0\ {\rm Mpc}^{-1}$ foreground modes in the raw data.

For the filter, we use a Gaussian process based filtering formalism from \citet{Kern2021} inspired by the DAYENU filtering method proposed by \citet{Ewall-Wice2020}, with the filter operator defined as
\begin{align}
\label{eq:gpr}
\bm{F} = \bm{I} - \bm{C}_{\rm fg}\left[\bm{C}_{\rm fg} + \bm{C}_{\rm noise}\right]^{-1},
\end{align}
where the foreground covariance $\bm{C}_{\rm fg}$ is taken to be a Sinc function (i.e. a tophat in delay space) with a rejection bandpass of $|\tau^{\rm max}| = 250$ nanoseconds, and the noise covariance is diagonal with a variance that is $10^{-8}$ times the foreground covariance amplitude.
Note that the above filter is mathematically equivalent to the DAYENU filter but with a different filter width.
The sharp delay filter suppresses power below $|\tau| <$ 250 ns ($|k_\para|\lesssim0.1$ Mpc$^{-1}$ for $z=10.4$) in all visibilities for all baselines, regardless of their length or orientation.
This filtering will also have interesting consquences for the response of the data to the sky brightness distribution.
In particular, it will downweight sensitivity to foreground emission near the peak response of the primary beam, thereby upweighting the relative importance of foreground emission coming from the observer's horizon.
We also validate the impact this filter has on the recovered 21\,cm power spectrum in \autoref{sec:HI_signal}.

\subsection{The 21\,cm Foreground Problem}
\label{sec:fg_problem}

The fundamental challenge of 21\,cm cosmology is in separating bright contaminating foreground emission from the background cosmological signal.
What makes this particularly difficult is the fact that foreground emission is thought to be $\sim10^5$ times brighter than the background signal,\footnote{This exact number depends on observing field, observing frequency, and the cosmological Fourier modes being probed, but is a good first-order estimate.} setting up an extremely delicate signal separation problem.
See \citet{Liu2020} and references therein for a review of the expansive foreground-modeling and subtraction studies for 21\,cm cosmology.
In effect, this places a requirement that foregrounds and spurious instrumental systematics be isolated to roughly 1 part in $10^5$.
This is a daunting challenge that has required new developments in radio data analysis methodologies, and has thus far precluded direct detection of the 21\,cm cosmological signal.


However, works studying the nature of smooth-spectrum foreground emission in interferometric datasets, like that generated by non-thermal synchrotron processes, have shown that foreground emission largely contaminates a wedge-like region of data in 2D Fourier space that can be identified and excised, known as the \emph{foreground wedge} \citep{Morales2004, Datta2010, Morales2012, Trott2012, Vedantham2012, Liu2014a}.
Taking the Fourier transform of the visibilities across frequency transforms them into \emph{delay space} ($\tau$),
\begin{align}
\label{eq:delay_transform}
\tilde{V}(\tau) = \int V(\nu) e^{-2\pi i \nu\tau} d\nu,
\end{align}
defined here such that the inverse transform picks up a normalizing $2\pi$.
The Fourier-transformed visibility is a means of directly accessing the 21\,cm power spectrum without having to make deep images of the sky, whose square is known as the delay power spectrum estimator \citep{Parsons2012b, Liu2014a, Thyagarajan2015a}.
\citet{Parsons2012b} showed that the delay spectrum can directly probe a windowed version of the power spectrum, where the delay and baseline length of the visibilities translate to the line-of-sight Fourier wavemode and transverse Fourier wavevectors:
\begin{align}
\label{eq:k_para}
k_\para &= \tau\ \frac{2\pi \nu_{21}H_0E(z)}{c(1+z)^2}, \\
\bm{k}_{\perp} &= \bm{b}\ \frac{2\pi}{\lambda D(z)}, 
\end{align}
where $\lambda$ is the central wavelength of the observing band, $\nu_{21}$ is the restframe 21\,cm transition frequency, $D(z)$ is the transverse comoving distance, $H_0$ is the Hubble constant, and $E(z) = [\Omega_m(1+z)^3 +\Omega_\Lambda]^{1/2}$ \citep{Liu2014a}.

If we assume the sky and the instrument to be frequency independent, for the moment, and we insert the complex exponential term from \autoref{eq:rime_integral} into \autoref{eq:delay_transform}, we see that it acts as a delta function in the delay transform, pushing the intrinsically $\tau\sim0$ foreground response to higher delays.
The extent of this effect is determined by the dot product $\tau_{pq}=\bm{b}_{pq}\hat{\bs}/c$, which achieves a maximum when radiation is incident from the observer's horizon: $\tau_{pq}^{\rm horizon} = b_{pq}/c$, which translates to $k_\para^{\rm horizon}$ via \autoref{eq:k_para}.
This means that, in principle, smooth-spectrum foregrounds should occupy a region between $\pm k_\para^{\rm horizon}$ in the Fourier-transformed visibilities.
This forms the basis for the ``foreground avoidance'' approach, which applies a high-pass filter to the visibilities that rejects signals in this region, resulting in residual $|k_\para| > k_\para^{\rm horizon}$ modes that are assumed to be foreground free.
However, in reality this is not the full story, as any \emph{additional} spectral structure from the instrument (say from the primary beam response or other instrumental effects), push foreground power to even higher delays, creating what is known as \emph{foreground leakage}.
Indeed, foreground leakage has been observed in most 21\,cm experimental results \citep{Pober2013b, Kern2020a, Mertens2020, Kolopanis2023}, and can be attributed to a variety of factors.

Thus we are left with a difficult question: at what point might we confuse foreground leakage with the real 21\,cm cosmological signal?
The natural question to ask is whether we can jointly model the complex interplay between foregrounds, instrumental effects, and the cosmological signal in order to 1) more robustly separate 21\,cm signal from systematics and 2) faithfully propagate covariant uncertainties from our foreground and instrumental models onto our 21\,cm signal reconstruction (i.e. marginalize the posterior distribution across our foreground and instrumental parameters).
Furthermore, we must be able to do this to very high precision given the large dynamic range between the contaminants and the cosmological signal.
This is the fundamental aim for an end-to-end model that can jointly explore foreground, instrumental, and signal parameters.

\subsection{The Posterior Probability Distribution}
\label{sec:posterior}

Let the parameters of our forward model (instrumental, foreground, and 21\,cm signal) be collected into a single column vector $\btheta$.
Given a choice of model parameters, we can \emph{simulate} the radio visibilities via a forward pass of our model (\autoref{eq:rime_matrix} \& \autoref{eq:rime_filter}), creating a set of model visibilities ($\bmm$) as a function of observing frequency, observing time, and baseline vector. When comparing these to raw data from a telescope ($\bd$), we need to write down a likelihood.
A Gaussian likelihood for the our data is
\begin{align}
\label{eq:likelihood}
\mathcal{L}(\bd|\bmm,\btheta) = \frac{\exp\left[-\tfrac{1}{2}(\bd-\bmm_{\btheta})^\dagger\bSigma^{-1}(\bd-\bmm_{\btheta})\right]}{(2\pi)^{n/2}|\bSigma|^{1/2}},
\end{align}
where $n$ is the dimensionality of the data, $\bSigma$ is the covariance matrix of the residuals, and $\bd$ and $\bmm$ are column vectors holding the data visibilities and model visibilities, respectively.
Noise in the raw data is well-modeled as Gaussian, however, the signal itself may have both Gaussian and non-Gaussian contributions. We defer exploration of non-Gaussian likelihoods and likelihood-free inference to future work. 
Given this, our adopted covariance matrix is populated with the noise variance along its diagional.


With the data likelihood in hand, we are prepared to make an inference of the model parameters by constructing the posterior probability distribution, or the probability density of the parameters given the data.
This is given by Bayes' theorem, which states that
\begin{align}
\label{eq:posterior}
P(\btheta|\bd) = \frac{\mathcal{L}(\bd|\bmm, \btheta)\pi(\btheta)}{\mathcal{Z}(\bd)},
\end{align}
where $P(\btheta|\bd)$ is the posterior distribution of the model parameters given the data, $\pi(\btheta)$ is the prior distribution of the parameters, and $\mathcal{Z}(\bd)$ is the marginal likelihood of the data, also known as the Bayesian evidence. The marginal likelihood acts as a normalization coefficient of the posterior, and is not strictly needed for parameter inference and credible interval calculation; however, it is often used for performing model selection, which we will defer to future work given its complexity.
The prior is critically important, and one of the advantages of the Bayesian approach is the ability to incorporate physically-motivated priors that can help steer inference.
This could be prior information about the foregrounds (say from first-principles arguments or from sky maps of other experiments), as well as prior information about the instrument itself (say from theoretical modeling or from lab measurements of the instrumental response).
We discuss our choice of priors for our proof-of-concept demonstration in \autoref{sec:parameterization}.
Note that for the optimization and sampling work described throughout the text, we will technically extremize the negative log posterior instead of the posterior itself, or $\mathcal{P}(\btheta|\bd)=-\log P(\btheta|\bd)$.

The complexity of the forward model makes navigating the posterior distribution difficult.
Depending on how we parameterize the signal, foregrounds, and systematics, the posterior can be poorly-conditioned and even degenerate.
However, this is not necessarily a deficiency of the end-to-end approach adopted here; rather, it is a statement on the reality of the difficult signal separation problem facing 21\,cm cosmology, where the desired signal is masked by foregrounds and systematics that can be partially degenerate with it.
Tools that enable us to fully explore these degeneracies, such as the one proposed in this work, are therefore critical.

\begin{figure*}
\centering
\includegraphics[width=\linewidth]{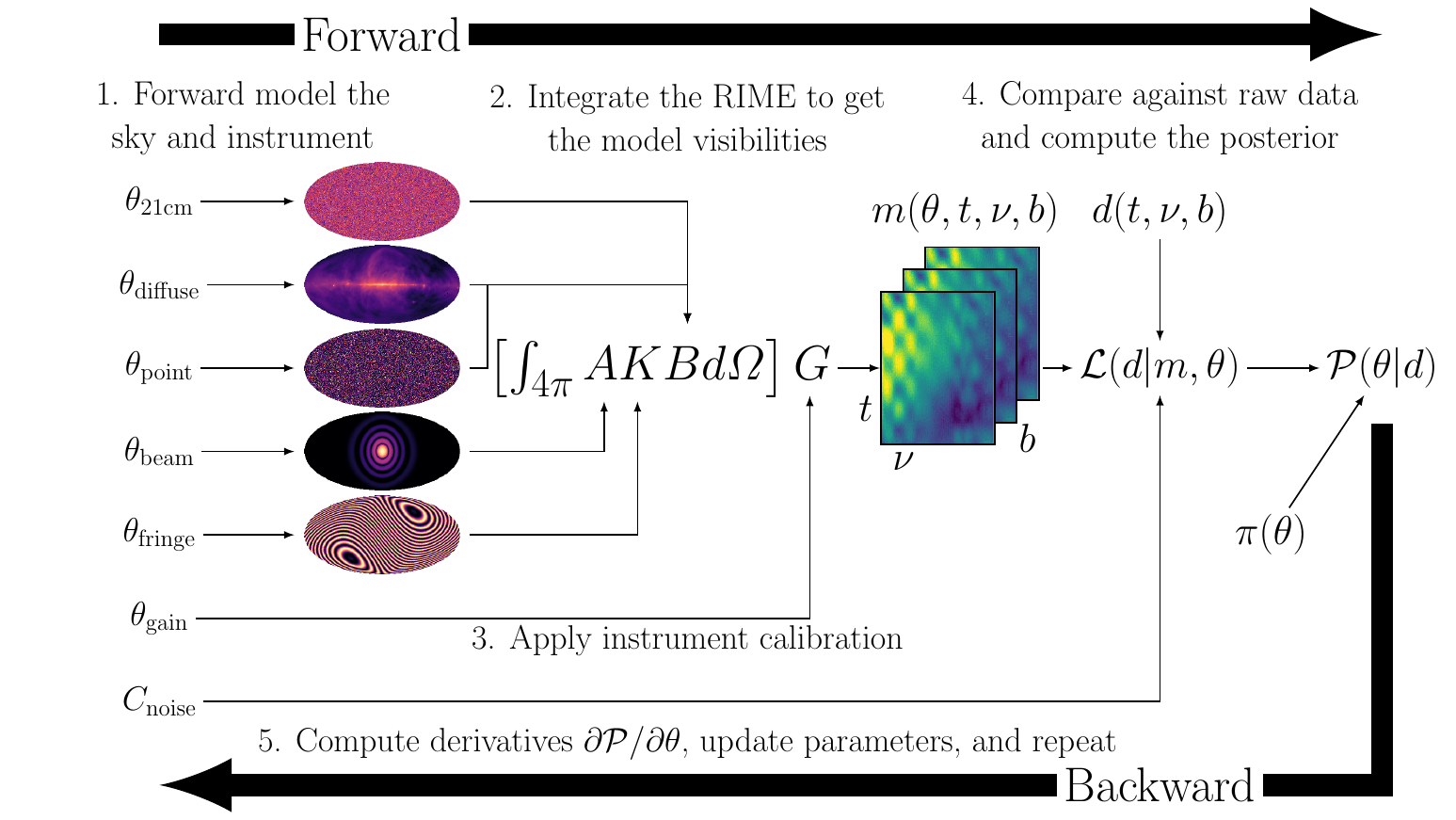}
\caption{A flowchart describing a Bayesian RIME forward model, starting with the model parameters (left column) and ending with the posterior probability of the parameters given the observed data (right). These parameters could include, for example, the 21\,cm signal ($\theta_{\rm 21cm}$), diffuse foregrounds ($\theta_{\rm diffuse}$), point source foregrounds ($\theta_{\rm point}$), antenna beam responses ($\theta_{\rm beam}$), antenna separation vectors ($\theta_{\rm fringe}$), direction-independent antenna gains ($\theta_{\rm gain}$), and a noise covariance of the data ($C_{\rm noise}$). Note that in this work we only treat $\theta_{\rm 21cm}$, $\theta_{\rm diffuse}$, and $\theta_{\rm beam}$ as free parameters, and treat $\theta_{\rm fringe}$, $\theta_{\rm gain}$ as both fixed and known.
Wrapping this forward model within an automatic differentiation engine allows us to compute gradients of the posterior with respect to the model parameters, which are then used for efficient parameter optimization and sampling.
}
\label{fig:flowchart}
\end{figure*}

To optimize such a posterior distribution and derive our best-fitting combination of model parameters, we need to compute the derivative of the posterior with respect to our model parameters.
Our approach for doing this leverages automatic differentiation (AD), specifically reverse-mode AD, which builds a computational graph of our forward model and then ``backpropagates'' through it to yield the desired gradients.
Reverse-mode AD applied to neural network models is also known as the backpropagation algorithm \citep{Baydin2015}, although it can be applied to models without neural connections all the same.
Indeed, our approach is to write a standard physical simulation with an AD-enabled backend to be able to leverage highly efficient gradient-based optimizers and samplers, which is a practice sometimes referred to as ``differentiable programming.''
Unlike finite-difference methods, automatic differentiation gradients are (numerically) exact, and are generally much faster to compute.
A flowchart of our end-to-end Bayesian forward model for 21\,cm cosmology is shown in \autoref{fig:flowchart}, which demonstrates how we simulate visibilities given a set of model parameters, compute a posterior, and then leverage AD to compute the gradient of the posterior with respect to our model parameters.
Note that \autoref{fig:flowchart} includes terms like antenna gain terms that are not explored in this work but are supported by \texttt{BayesLIM}.
Our framework is built on \texttt{PyTorch} \citep{pytorch} and uses its reverse-mode automatic differentiation library.


\section{Model Parameterization}
\label{sec:parameterization}
Here we discuss our choice of parameterization for the components in our forward model, as well as the specifications for our mock HERA observations.
In what follows, we will specify a data model for 21\,cm intensity mapping at EoR redshifts, however, note that many of these parameterization choices are equally valid for low redshift 21\,cm intensity mapping as well.
Furthermore, the exact choice of parameterization may be context-dependent, and the process of selecting an optimal parameterization for a given telescope design is still an area of study.
Also note that the process of model selection, or determining the degrees of freedom of the model, is a critical question that can be addressed by computing the Bayesian evidence factor in \autoref{eq:posterior} \citep[e.g.][]{Sims2020, Murray2022}. However, this is computationally very expensive, particularly for the large number of parameters used in our forward model, and we defer exploration of this topic to future work.

As a concise summary, the parameters of our forward model that we actually optimize in this work include:
\begin{enumerate}
    \item[1.] \emph{Antenna Primary Beam Response} -- We model the anntena primary beam total power response pattern (assumed to be shared by all antennas) with 75 (real-valued) spherical harmonic angular modes ($\ell_{\rm max} = 40$ and $m_{\rm max} = 6$) and 5 orthogonal polynomial modes across frequency, for a total of 385 parameters. We set a Gaussian prior on the beam in real space centered at the fiducial model, with a variance that yields $1\sigma$ beam fluctuations at roughly -25 dB, which is generally consistent with our prior knowledge of antenna primary beams \citep{Line2018, Nunhokee2020}.

    \item[2.] \emph{Foregrounds} -- We model the (diffuse + point source) foregrounds with \emph{spherical cap harmonic modes} (discussed below) up to $\ell_{\rm max}=160$, which covers the full horizon-to-horizon observable sky given HERA's observing coordinates. We use 12,104 harmonic angular modes and 3 orthogonal Legendre polynomials across frequency, for a total of 36,312 (complex-valued) parameters. We adopt a Gaussian prior on the spherical harmonic coefficients that translates to a $\sim5\%$ uncertainty on the starting fiducial foreground map, which is roughly consistent with our current understanding of the low-frequency foreground distribution \citep{Zheng2017}.
 
    \item[3.] \emph{EoR Signal} -- We model the EoR signal with \emph{spherical stripe harmonic modes} (discussed below) out to the same angular resolution as the foreground model ($\ell_{\rm max}=160$), which covers $\sim4000$ square degrees across a drift-scan observing mask tracking the main field-of-view of the simulated HERA observations. We use 1,302 harmonic modes to model the angular dimension and 40 orthogonal polynomials for the frequency dimension for a total of 52,080 complex-valued parameters. We set a weak, mean-zero Gaussian prior on the harmonic coefficients, with a variance that is ten times greater than the variance of the mock 21\,cm model used as the true underlying signal. This is meant to act as a minimally informative prior model, while still regularizing the modes to prevent them from taking on unrealistic values that would exceed current upper limits.
\end{enumerate}
In total, our forward model contains roughly $\sim88,000$ complex-valued parameters across the instrument, foreground, and 21\,cm signal components.

\subsection{Array Model and Mock Observations}
\label{sec:array}

\begin{figure}
\centering
\includegraphics[width=.9\linewidth]{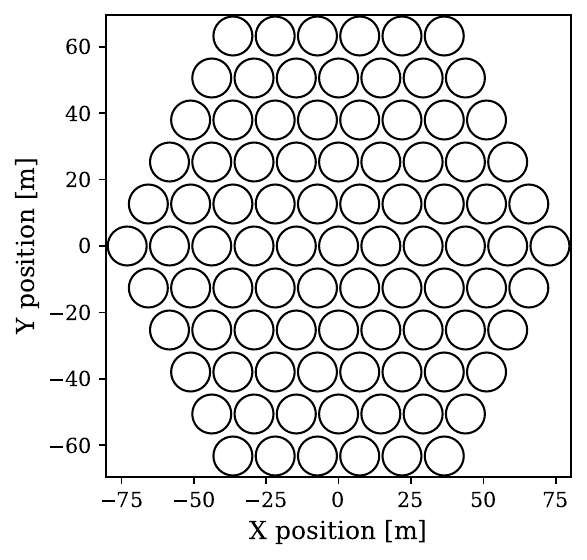}
\caption{The modified HERA-91 array layout adopted in this work. The array consists of 91 antennas with 14.6-meter spacing. Note that in this work we only simulate baselines with lengths from $0 < |\bm{b}| < 60$ meters due to constraints on the angular resolution of our adopted sky model; however, this baseline cut still preserves nearly 80\% of the total power spectrum sensitivity of the array after foreground wedge filtering.}
\label{fig:hera_array}
\end{figure}

We use a condensed version of the HERA array as a prototype for testing our framework, shown in \autoref{fig:hera_array}. This consists of 91 antennas packed in a hexagonal fashion with 14.6 meter spacing between antennas, similar to the HERA design without the split-core feature \citep{DeBoer2017}.
For this proof-of-concept study, we will only analyze data from baselines with lengths between $0<|\bm{b}|<60$ meters, thus excluding the auto-correlation visibilities ($|\bm{b}|=0)$ and the long baseline visibilities.
The baseline cut is mainly for computational reasons due to the limited angular resolution of our foreground model; however, even with this baseline cut we preserve nearly 80\% of the array's power spectrum sensitivity between $0 < k < 0.35\ h\ {\rm Mpc}^{-1}$, assuming we've applied a horizon-wedge FG filter that is similar in specification to the \emph{pessimistic} foreground case in \citet{Pober2014}. This leaves a total of 30 unique baseline vectors that we simulate via \autoref{eq:rime_integral}, which are then broadcasted to 1,785 physical baselines that are used as the model visibilities. This distribution of baseline lengths (combined with the frequency band described below) cover transverse Fourier modes between $0.004\le |k_\perp| \le 0.016\ {\rm Mpc}^{-1}$.

Our simulated frequencies span a 10 MHz bandwidth from 120 -- 130 MHz, yielding a central redshift of $z\sim10.4$ for the 21\,cm line. This aligns with one of the main cosmology observing bands in \citet{HERA2022a}.
We simulate the data with a spectral resolution of roughly 222 kHz, which is somewhat more coarse-grain than typical 21\,cm experiments; however, in this study we are mainly aiming to recover intermediate $k_{\para}$ modes, largely because the high $k_{\para}$ modes of most EoR models (i.e. $|k|>1\ {\rm Mpc}^{-1}$) are nearly entirely noise dominated, even for second-generation 21\,cm experiments. A 10 MHz bandwidth with 222 kHz spectral resolution allows us to model cosmological line-of-sight Fourier modes between $0 \le |k_\para| \le 0.75\ {\rm Mpc}^{-1}$, however, as noted above, we employ a frequency-based high-pass filter that eliminates power in the visibilities for $|k_\para| \le 0.1\ {\rm Mpc}^{-1}$

Lastly, our mock HERA dataset is simulated along a contiguous 6-hour drift scan from a local sidereal time (LST) of $0 < t < 6$ hours, which tracks right ascensions of $0 < \alpha < 90$ degrees at a fixed declination of -30.72 degrees. This LST range covers the main fields-of-interest used in previous HERA results \citep{HERA2022a, HERA2023}. We simulate 68 time integrations evenly spaced throughout the 6-hour interval, resulting in a time difference of 5 minutes between distinct snapshot observations. While this is much longer than a typical observing cadence of real HERA data, on the order of tens of seconds, it is still below HERA's beam-crossing time of roughly 30-minutes (the time it takes a celestial source to transit the antenna primary beam).
This allows us to effectively interpolate between time integrations without significant loss of signal if needed. Also note that the final time binning cadence in recent HERA results (after calibration) are on the order of 5 minutes \citep{HERA2022a}.
\autoref{fig:sky_sampling} shows the sky regions used for modeling the foreground and EoR sky signals (top), showing how the diffuse model covers the entire observable sky from HERA's coordinates, while the EoR model need only cover the main FoV of the drift-scan observations. It also shows the maximum primary beam response throughout the drift-scan observations (bottom), demonstrating that while most of the telescope's sensitivity is contained within a stripe at fixed declination, the full observable sky is still measured at attenuations of $10^{-3}-10^{-4}$, which is enough to allow bright, off-axis foregrounds like the galactic plane to dominate the intrinsic EoR 21\,cm amplitude in the visibilities.

\subsection{Foreground Model}
\label{sec:foregrounds}

The dominant form of unpolarized astrophysical foregrounds come from non-thermal synchrotron radio emission in the galaxy and from extragalactic sources.
Synchrotron continuum flux density follows a powerlaw form of $S\propto\nu^{-\alpha}$ with a spectral index of $\alpha\sim2.2$ \citep{Condon1992, Haslam1982, Remazailles2012}.
As a consequence of the power-law form, these foregrounds are particularly bright at the low radio frequencies used for 21\,cm cosmology measurements, reaching up to $10^5$ times brigher than the expected 21\,cm cosmological signal.
A blessing of the power-law form, as discussed previously, is the assumed smoothness of the continuum as a function of frequency.
However, the angular distribution of the foregrounds is more complex.

Considerable effort has gone into improving our understanding of these foregrounds for 21\,cm cosmology science, particularly at the relatively less-studied frequency bands below 1000 MHz.
This includes surveys of the vast population of radio point sources \citep[e.g.][]{Cohen2007, Hurley-Walker2017, Riseley2020, Hurley-Walker2022}, surveys of diffuse emission from the galaxy \citep[e.g.][]{Haslam1982, Oliveira2008, Remazailles2012, Zheng2017, Dowell2017, Eastwood2018, Mozden2019, Spinelli2021, Byrne2022} and their polarized structures \citep{Jelic2010, Moore2013, Nunhokee2017}, and studies of individual, nearby resolved radio galaxies, like Fornax A \citep{McKinley2015, Line2020}.
These synergies have been highly beneficial to the field as a whole, a recent example being how the HERA experiment leveraged the GLEAM survey as a key component in its absolute calibration pipeline \citep{Kern2020b}.

Recently it has become increasingly clear that robust foreground modeling requires a model of the full sky, as opposed to simply the main field-of-view \citep{Pober2016, Bassett2021}. In particular, diffuse foregrounds near the observer's horizon creates the now well-studied phenomenon known as the \emph{pitchfork effect} \citep{Thyagarajan2015a}, which has been observed in simulations \citep{Kern2019, Lanman2020, Charles2023} and in the raw data of a variety of 21\,cm telescopes \citep{Thyagarajan2015b, Kern2020a, Rath2024}.
The pitchfork effect is particularly troublesome because the foregrounds manifest in the visibilities on the boundary of the foreground wedge at $|\tau_{\rm horizon}|$, and are easily leaked into the EoR window from instrumental chromaticity.

\begin{figure}
\centering
\includegraphics[width=\linewidth]{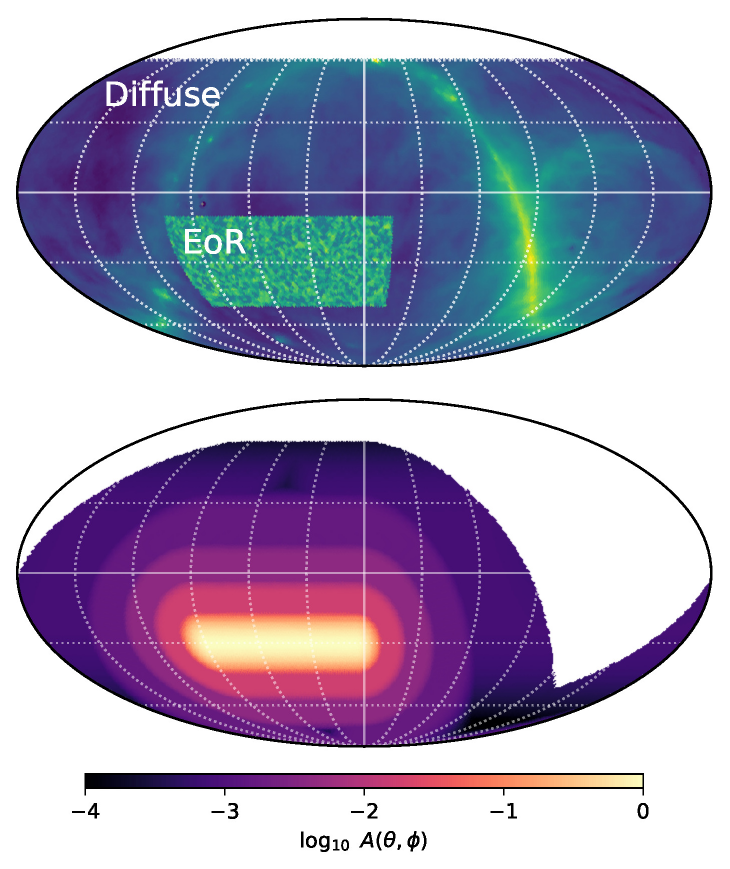}
\caption{Top: We plot the angular coverage of our foreground model, which spans the full observable sky from HERA's coordinates (diffuse). We also show the EoR sky model coverage, which tracks a smaller ``stripe'' across 120 degrees in right ascension at fixed declination. The EoR model amplitude is artificially boosted for visual clarity. Bottom: The maximum primary beam amplitude across the full 6 hour simulated drift-scan observation. This shows the maximum response of the telescope at each sky pixel over the course of the observations, showing that while most of HERA's sensitivity is confined within a narrow stripe on the sky, it is still sensitive to bright, off-axis foregrounds at an attenuation of $\sim10^{-3}-10^{-4}$, which is enough to dominate the intrinsic EoR signal in the simulated visibilities.}
\label{fig:sky_sampling}
\end{figure}

Our foreground model therefore spans the entire observable sky from HERA's coordinates (\autoref{fig:sky_sampling}).
We start with a fiducial model of the diffuse sky from the Global Sky Model catalogue \citep{Haslam1982, Oliveira2008, Remazailles2012, Zheng2017}, specifically the updated 2016 model \citep{pygdsm}, which combines low-frequency measurements of the sky into a series of best-understanding maps at our observing frequencies.
Note that while some care has gone into removing different telescope artifacts and bright, extended radio sources in constructing the GSM \citep{Remazailles2012}, the model still contains a background extragalactic point source distribution.

After evaluating the GSM at each of our observing frequencies, we interpolate the foreground map onto an fixed, equal-area rectangular grid in right ascension and declination with an effective cell resolution of 0.5 degrees, which is similar to a HEALPix NSIDE 128 resolution.
This converts the continuous foreground sky brightness distribution with units of specific intensity into pixelized cells with units of flux density, specifically Jansky.
This is akin to our RIME integral pixelization in \autoref{eq:rime_matrix}, with the grid extending over the entire observable sky from HERA's coordinates (\autoref{fig:sky_sampling}).
We tested the accuracy of this 0.5 degree pixelization for the telescope setup described in \autoref{sec:array}, and found accurate reconstruction of a higher resolution HEALpix NSIDE 256 discretized simulation with a fractional RMS of $\sim10^{-5}$.
Although this error level is right on par with the FG-to-EoR dynamic range tolerance, we can get away with this because what we really care about are the FG modes that fall near the boundary of the FG wedge, which for HERA is around $k\sim0.1\ {\rm Mpc}^{-1}$.
These modes are also naturally attenuated by the beam, and therefore have a lower FG-to-EoR dynamic range, closer to $10^3-10^4$.
It is true that certain systematics, like very poorly behaved antenna gains, can significantly leak the $k=0\ {\rm Mpc}^{-1}$ modes outside the FG wedge, but we do not explore these effects in this work.
Future work will study the necessity of higher sky resolutions when working with more complex systematic models.

We parameterize the angular response of the foregrounds in a spherical harmonic basis, using the spherical cap harmonic formalism (see \autoref{sec:ssfb}).
Briefly, the spherical cap harmonics \citep{Haines1985} are a modified spherical harmonic basis that is complete and orthogonal on the spherical cap (as oppposed to the full sphere).
This allows for a compressed basis for modeling signals on the cut sky, with the tradeoff being non-integer-valued $\ell$ modes.
The forward transform of the angular coefficients into map space is defined as
\begin{align}
\label{eq:fg_harmonics}
B^{\rm fg}(\hat{s},\nu) = \left|{\rm Re}\left(\sum_{\ell m}^{\ell_{\rm max}}Y_{\ell m}^{\rm fg}(\hat{s})a_{\ell m}^{\rm fg}(\nu)\right)\right|,
\end{align}
where $B^{\rm fg}(\hat{s},\nu)$ is the real-valued, non-negative flux density of the pixelized foreground sky, $Y_{\ell m}^{\rm fg}$ are the complex-valued spherical cap harmonics as a function of sky angle, $a_{\ell m}^{\rm fg}(\nu)$ are the spherical cap harmonic coefficients as a function of frequency channel, and the sum runs over all $\ell$ and $m$ modes up to $\ell_{\rm max}$.
Note that due to the real-valued nature of the unpolarized foreground sky, we can throw out all negative $m$ modes in $Y_{\ell m}$ and simply multiply the $m>0$ fitted coefficients by a factor of two when taking the forward transform.
In matrix form, we can solve for the best-fit harmonic coefficients given a map of the foreground sky via their least squares solution, given as
\begin{align}
\label{eq:fg_harmonic_matrix}
\hat{\bm{a}}^{\rm fg} = (\bm{Y}^T \bm{Y})^{-1} \bm{Y}^T \bm{B}^{\rm GSM},
\end{align}
where $\bm{B}^{\rm GSM}$ is the matrix of GSM foreground maps in $\mathbb{R}^{N_{\rm pix}\times N_{\rm \nu}}$, $\bm{Y}$ is a matrix of spherical cap harmonic modes in $\mathbb{C}^{N_{\rm pix}\times N_{\rm modes}}$ and $\hat{\bm{a}}$ are the best-fit harmonic coefficients in $\mathbb{C}^{N_{\rm modes}\times N_\nu}$.
We model all $\ell\ \&\ m$ modes up to an $\ell_{\rm max}=160$ cutoff for a total of 12,104 coefficients, beyond which the telescope is not particularly sensitive given the maximum baseline in our data and the observing frequencies.\footnote{Convergence tests show we can recover the foreground power in the longest baselines with a fractional error of $\sim10^{-4}$ with the selected $\ell_{\rm max}=160$ relative to an unsmoothed foreground map.}
Note that the choice of an effectively band-limited model of the diffuse sky given the angular resolution of the telescope means that the foreground model acts effectively as a joint diffuse and point source model.
The desire to have a band-limited foreground model in this work is what drives the maximum baseline cutoff of 60 meters, beyond which the number of foreground parameters becomes cumbersome to work with (but perhaps not technically computationally infeasible).
Future work will explore other angular parameterizations that may enable a higher $\ell_{\rm max}$ cutoff.

The frequency axis is parameterized with a second-order orthogonal Legendre polynomial (3 coefficients) that enables recovery of the GSM powerlaw-like structures down to a fractional error of $10^{-5}$ over our 10 MHz observing bandwidth.
The forward transform from the polynomial coefficient domain into the frequency domain is given as
\begin{align}
\label{eq:fg_legendre}
a_{\ell m}^{\rm fg}(\nu) = \sum_{k=0}^2 X_k^{\rm fg}(\nu)\ \tilde{a}_{\ell mk}^{\rm fg},
\end{align}
where $X^{\rm fg}_k(\nu)$ are the foreground Legendre coefficients and $\tilde{a}_{\ell mk}$ are the fully compressed foreground parameters.
This leads to a total of 36,312 complex-valued parameters for our foreground model.

The native GSM model acts as our starting fiducial model of the low-frequency foreground sky.
Fitting the harmonic and Legendre modes to these multi-frequency maps creates our initial parameter vector, $[\tilde{\bm{a}}^{\rm fg}_{\ell mk}]_0$, which acts as our starting point before optimization.
To simulate a mock HERA observation we perturb the model about this starting point to act as a pseudo ``ground truth'' that is assumed to be a priori unknown.
We do this by adding random Gaussian noise to the fiducial set of coefficients via
\begin{align}
[\tilde{\bm{a}}_{\ell mk}^{\rm fg}]_{\rm truth} = [\tilde{\bm{a}}_{\ell mk}^{\rm fg}]_0 + \bm{n}_{\ell mk},
\end{align}
where $\bm{n}_{\ell mk}\sim\mathcal{N}(0, [\sigma_{\rm prior}^{\rm fg}]^2)$.
We tune the amplitude of the noise such that it yields a forward modeled foreground map that has a residual fractional standard deviation that is $\sim5\%$ of the fiducial foreground map amplitude, which is roughly consistent with our current understanding of low-frequency foregrounds \citep{Zheng2017}.
In other words, we tune the noise amplitude $\sigma_{\rm prior}^{\rm fg}$ such that the standard deviation of the ratio $B_{\rm truth}^{\rm fg}/B_{\rm fiducial}^{\rm fg}$ is roughly 0.05.
Finally, we set a Gaussian prior directly on the harmonic coefficients with a mean of $[\tilde{\bm{a}}_{\ell mk}^{\rm fg}]_0$ and a diagonal covariance matrix with a scalar amplitude of $[\sigma_{\rm prior}^{\rm fg}]^2$.

Lastly, we have a few final notes about our foreground parameterization for the avid practitioner.
In particular, the bottleneck in the foreground forward model transform is the angular transform by the spherical cap harmonic matrix $\bm{Y}$, which for the specifications listed above would make it a 36,000 $\times$ 153,360 matrix, requiring 45 GB of RAM to store in computer memory (assuming a double precision, complex floating point data type).
This is particularly cumbersome when running GPU-accelerated automatic differentiation as the matrix needs to be stored in GPU memory, which is significantly more limited compared to a generic CPU cluster.
However, one can significantly decrease the size of this matrix by making it separable along the right ascension and declination axes, which is possible if we choose a uniform, rectangular grid sampling.
In this case, the foreground forward transform can be written as
\begin{align}
\label{eq:fg_separable}
B^{\rm fg}(\hat{s},\nu) = \left|{\rm Re}\left(\sum_{\ell m}\Theta_{\ell m}^{\rm fg}(\theta)\Phi_m^{\rm fg}(\phi)\sum_{k}X_k^{\rm fg}(\nu)\tilde{a}_{\ell m k}^{\rm fg}\right)\right|,
\end{align}
where $\theta$ and $\phi$ are spherical polar and azimuthal angles, respectively, $\Theta_{\ell m}(\theta)$ are the Associated Legendre polynomials, and $\Phi_m(\phi)$ are the standard Fourier series (see \autoref{sec:ssfb} for more details).
Having made the forward transform separable onto a rectangular grid of points in $(\theta, \phi)$, we need only store $\Theta_{\ell m}(\theta)$ of size $N_{\theta}$ and $\Phi_m(\phi)$ of size $N_\phi$, which are both considerably smaller than the $N_\theta\times N_\phi$ dimensionality of $Y_{\ell m}^{\rm fg}(\theta,\phi)$.

\subsection{Cosmological 21 cm Signal Model}
\label{sec:HI_signal}

To model the cosmological 21\,cm signal from the EoR, we first run a semi-numerical simulation of the 21\,cm differential brightness temperature signal $\delta T_{21}$ using the \texttt{21cmFAST} code \citep{Mesinger2011}.
We use the default astrophysical and cosmology parameters found in the version-2 code (\url{https://github.com/andreimesinger/21cmFAST}), which puts the 21\,cm reionization history in agreement with existing probes at the end of reionization \citep[e.g.][]{Park2019}.
We simulate a volume with side length $L=800$ Mpc and periodic boundary conditions with a cell resolution of 1 Mpc.
Next, we band-limit the simulations by applying a Sinc filter that removes signal above $k\sim0.75\ {\rm Mpc}^{-1}$, which is the largest Fourier wavemode probed given our frequency channelization.
Next we use the ``tile-and-interpolate'' procedure of converting the simulation box output onto a series of full-sky maps \citep{Kittiwisit2022}.
To do this, we tile the 21\,cm simulation box in 3D space out to the line-of-sight comoving distance of our observations between $10 < z < 10.8$, and peform nearest neighbor interpolation onto a high resolution HEALpix grid of NSIDE 2048.
We then apply a smoothing filter to bandlimit the maps before interpolating onto the 0.5-degree rectangular grid used for the foreground model.
However, in this case, we only use a small cutout of the main field-of-view of the HERA primary beam response instead of using the full observable sky (see \autoref{fig:sky_sampling}).
This spherical stripe region spans 120 degrees in right ascension and 40 degrees in declination, covering the sky where the primary beam remains over 1\% of its peak value throughout the drift-scan observations.
We need only model the EoR signal over this smaller area because the vast majority of the cosmological signal enters through this region of the sky due to primary beam attenuation in the far sidelobes.

\begin{figure}
\includegraphics[width=\linewidth]{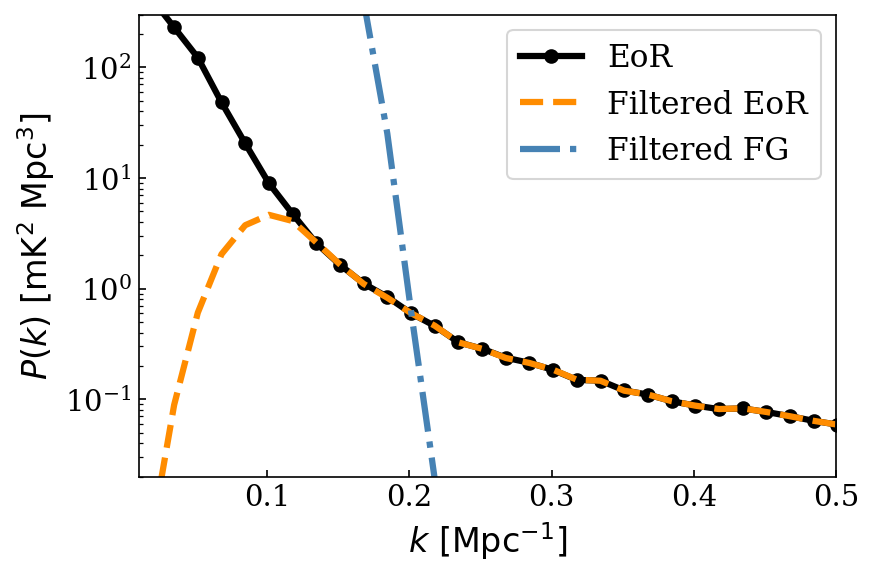}
\caption{Comparison of a 21\,cm power spectrum produced from forward-modeled interferometric visibilities of an EoR sky model (black), and a power spectrum produced from the same visibilities having applied the delay filter (orange-dashed) from \autoref{eq:rime_filter}. The filter suppresses signal in the power spectrum for $k\le0.1\ {\rm Mpc}^{-1}$ and leaves other Fourier modes intact. We also plot a power spectrum from the (delay filtered) foreground component of our forward model to demonstrate the range of Fourier modes that would be contaminated without any sort of foreground subtraction, which includes effects from the instrumental response.}
\label{fig:filtered_eor}
\end{figure}

We parameterize the 21\,cm EoR sky signal with the spherical stripe harmonics (SSH), introduced in \autoref{sec:ssfb}.
Like the spherical cap harmonics, the SSH are a modified version of the spherical harmonics that form a complete and orthogonal basis but for a spherical stripe geometry, sometimes also known as a spherical segment (see \autoref{fig:sky_sampling}).
This allows us to form a sparse basis given our observing mask while retaining certain statistical properties such as band-limited completeness.
We review the SSH and its 3D analog, the spherical stripe Fourier-Bessel formalism, in detail in \autoref{sec:ssfb}.
We model the EoR signal up the same $\ell_{\rm max}=160$ bandlimit of the foreground model, resulting in 1,302  complex-valued coefficients, significantly less than the $\sim$12,000 modes used for the full-sky foreground map with the same $\ell_{\rm max}$.

Like the foreground model, we decompose the harmonic transformation into separable polar and azimuthal transformations, while also using the same equal-area, 0.5 degree resolution sampling pattern as the foreground model.
However, unlike the foreground model, we do not limit the sky maps to be non-negative.
This is because we are modeling the differential brightness temperature, $\delta T_{21}$, relative to the CMB temperature.
Although the total sky brightness is still a non-negative quantity, in practice, $\delta T_{21}$ will never be negative enough to drive the total sky brightness to a negative quantity given our prior model.

For the frequency axis we also use a set of orthogonal Legendre polynomials similar to the foreground model, but now use 40 coefficients to be able to capture the fine frequency fluctuations found in the 21\,cm signal.
This leads to a total of 52,080 complex-valued parameters for the EoR component of the data model.
Thus, our full forward transformation from coefficient space to map space for the 21\,cm sky model is given as
\begin{align}
\label{eq:eor_model}
B^{21}(\hat{s}, \nu) = {\rm Re}\left(\sum_{\ell m}\Theta_{\ell m}^{\rm 21}(\theta)\Phi_m^{\rm 21}(\phi)\sum_k X_k^{\rm 21}(\nu)\tilde{a}_{\ell mk}^{\rm 21}\right).
\end{align}
The true coefficients for the 21\,cm mock observation, $[\tilde{\bm{a}}^{21}_{lmk}]_{\rm truth}$, are computed by fitting them to the simulated 21\,cm maps described above.
Because there are no direct constraints on the EoR 21\,cm field to date, our initial starting model for the 21\,cm field is taken to be a vector of zeros.
We set a weakly informative prior on the complex-valued 21\,cm harmonic parameters with a mean of zero and a variance that is ten times times greater than the variance of the fitted truth parameters.
This acts as a minimally informative prior model for the currently weakly constrainted 21\,cm field, while still regularizing them to prevent them from taking on unrealistic values that would exceed current upper limits on the signal.

In \autoref{fig:filtered_eor} we show the 21\,cm power spectrum generated by the described EoR model.
We also show the impact of the delay filter described in \autoref{sec:rime}, where to do so we have forward modeled the EoR sky model into a set of visibilities, applied the delay filter, and then estimated the power spectrum from the visibilities (discussed in \autoref{sec:ssfb_pspec}).
We see, as expected, a sharp cutoff in power at $k\le0.1\ {\rm Mpc}^{-1}$ for the filtered dataset, while other modes remain untouched.

\begin{figure*}
\centering
\includegraphics[width=\linewidth]{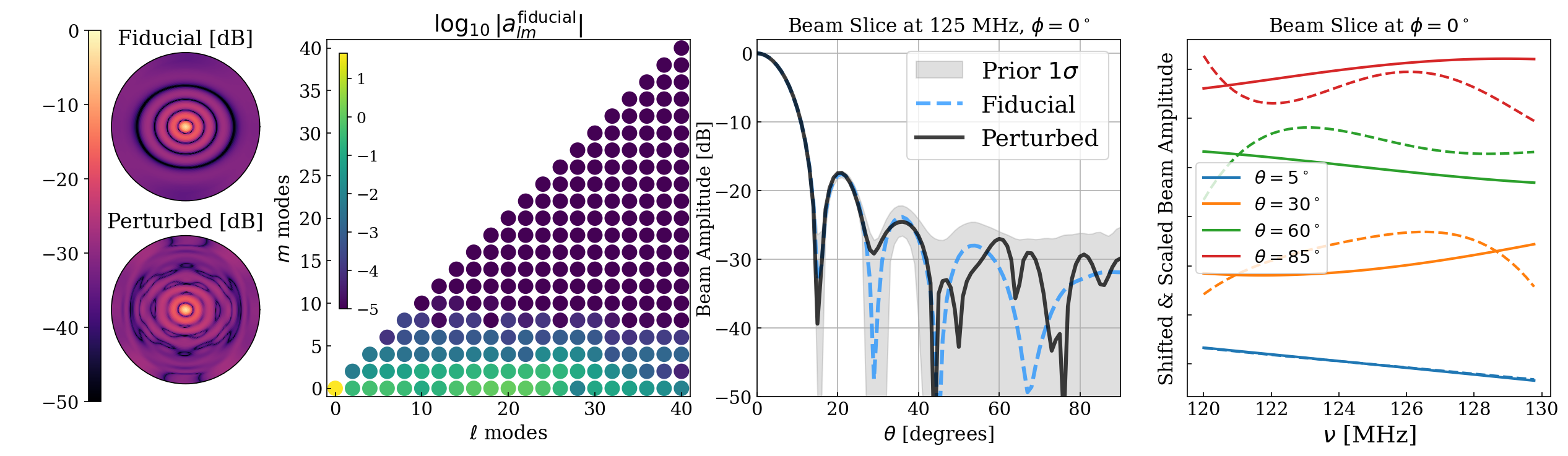}
\caption{
The fiducial and perturbed primary beam response as a function of angle and frequency.
{\bfseries Left}: Polar projection of the total power beam response at 125 MHz for the fiducial (top) and perturbed (bottom) beam in decibels.
{\bfseries Center-Left}: The log amplitude of the $a_{lm}$ decomposition of the fiducial beam, showing compression of the beam to low $m$ modes. The hemispherical cap harmonics allow even-valued sampling of $l$ modes, while the assumed 180$^\circ$ degree beam symmetry allows even-valued sampling of $m$ modes and the real-valued nature of the beam allows us to drop negative $m$ modes (they are just complex conjugates of positive $m$ modes), resulting in even further compression. Taking all $a_{lm}$ modes for $m\le6$ and $l\le40$ results in just 78 real-valued parameters for the beam's angular dimension. These 78 parameters fit the fiducial modified Airy pattern with a residual RMS less than $10^{-4}$.
{\bfseries Center-Right}: A slice through the beam amplitude of the fiducial beam (black) and the perturbed beam (red-dashed) in decibels, showing fairly complex structure in the perturbed beam, especially at large zenith angles. We also show the $1\sigma$ width of the prior (gray shaded) centered on the fiducial beam model.
{\bfseries Right}: A slice through the beam amplitude across frequency at fixed zenith angle (artifically normalized and offset for visual clarity), demonstrating the kind of non-trivial frequency structure found in the perturbed beam (dashed) relative to the fiducial beam (solid).
}
\label{fig:beam_decomp}
\end{figure*}

\subsection{Antenna Primary Beam Model}
\label{sec:beams}

The antenna primary beam response is one of the leading instrumental systematics for 21\,cm cosmology, and deserves particular attention \citep[e.g.][]{Shaw2014, Sokolowski2017, Tauscher2018, Line2018, Kim2023, Wilensky2024}.
Here we adopt a single model for all antennas (sometimes referred to as the ``average beam''), which has both angular and frequency degrees of freedom.

Our fiducial beam model is modeled as an Airy disk, which is a good first-order approximation of the HERA antenna response given that the dish carves out a circular aperture.
However, we make a slight modification to account for the natural squashing of the beam along the east or north direction (for the east or north-oriented feed, respectively) that arises from the response of the feed.
Our modified Airy disk function is written as
\begin{align}
\label{eq:airy}
A(\theta, \phi, D_{\rm ew}, D_{\rm ns}, \nu) &= 2 J_1(x)/x
\end{align}
where
\begin{align}
x &= [D_{\rm ns} + |\sin(\phi)|^2 (D_{\rm ew} - D_{\rm ns})]\sin(\theta)\pi\nu/c,
\end{align}
and $J_1$ is the Bessel function of the 1st kind of order 1.
Here, we replace the aperture diameter in the standard Airy disk function with an ``effective'' diameter that looks larger or smaller depending on the azimuth angle, creating the squashing effect.
The square of this function is used as a model of the total power of the antenna primary beam. We defer modeling the polarized primary beam reponse to future work.

While the modified Airy function represents our fiducial (or starting beam model), the ``truth'' beam model used in simulating our mock, raw dataset is a perturbation about this fiducial model.
To generate this perturbation, we decompose the beam model using the spherical cap harmonic formalism \citep{Haines1985}.
In our case, we assume the beam model response covers the full hemisphere above the observer horizon, with a $\theta_{\rm max}=90^\circ$.
In effect, this means that we use the standard spherical harmonic basis but truncate the odd $\ell$ modes. In the general case of any spherical cap (not just a hemispherical cap), this would translate to a new set of \emph{non-integer} $\ell$ modes, as is the case for the foreground model described above.
We describe the spherical cap harmonics and their associated spherical stripe harmonics in more detail in \autoref{sec:ssfb}.

We make one modification to the spherical cap harmonics to enable easier fitting to real beam data.
First, based on our definition of the primary beam in \autoref{eq:rime_integral}, the total power beam is a unitless quantity that is normalized such that the zenith pointing ($\theta=0^\circ$) should be equal to one.
However, all of the $m=0$ spherical harmonic modes have a non-zero response at $\theta=0^\circ$, meaning there is a tight degeneracy between these modes when fitting the beam near boresight.
We could set a very tight prior on our beam amplitude at $\theta=0^\circ$ to enforce this property, however, experimentation has shown this creates a posterior that is difficult optimize.
Instead, we reparameterize the $m=0$ modes by replacing the $\ell=0$ monopole mode with a Gaussian function that is fit to the envelope of the beam's main lobe. We then subtract this function from all other $m=0$ modes, such that all modes (except for $\ell=0$) go to zero for $\theta\rightarrow0^\circ$.
We then leave the $\ell=0$ mode fixed and only fit $\ell>0$ modes when optimizing for the beam shape.
The angular parameterization is therefore defined as,
\begin{align}
\label{eq:beam_harmonics}
A(\hat{s}, \nu) = \left|\sum_{lm}Y_{lm}^{\rm beam}(\hat{s})\  a_{lm}^{\rm beam}(\nu)\right|,
\end{align}
where $A(\hat{s})$ is the total power primary beam in \autoref{eq:rime_integral}.
We use an absolute value operator to enforce the intrinsic non-negativity of the total power beam.
One could also enforce this by modeling the log power beam, or by setting a non-negative prior on the angular representation of the beam. Based on experimentation, however, we found that taking the absolute value was the most efficient way to enforce this property without degrading the natural sparsity of the harmonic basis.

To model the frequency dependence of the beam we use a set of orthogonal polynomials defined across the observing bandwidth. Specifically, we use a 4th-order Legendre polynomial that is able to capture the intrinsic frequency structure of the fiducial Airy model down to a fractional RMS of $10^{-5}$.
Thus, we represent the frequency dimension of the fitted $a_{\ell m}^{\rm cap}$ harmonic modes as,
\begin{align}
\label{eq:beam_legendre}
a_{\ell m}^{\rm beam}(\nu) = \sum_k X_k^{\rm beam}(\nu)\ \tilde{a}_{\ell mk}^{\rm beam},
\end{align}
where $X_k^{\rm beam}(\nu)$ is the design matrix holding the 5 orthogonal Legendre polynomials in our 4th-order polynomial model, and $\tilde{a}_{\ell m}^{\rm beam}$ are the fully compressed modes of our frequency and angularly dependent primary beam model.

\autoref{fig:beam_decomp} shows the spherical harmonic decomposition of the this fiducial Airy model, showing good compression of the beam in harmonic space with $m \le 6$ and $\ell \le 40$. Furthermore, we achieve even further compression from the fact that we sample even-valued $\ell$ modes due to the hemispherical cap harmonics; we sample even-valued $m$ modes due to the assumed $180^\circ$ symmetry of the beam, and we sample only positive $m$ modes because the power beam is intrinsically real-valued. This results in the beam being well-compressed down to only 78 modes given the $\ell m$ cuts described above, which we find can represent the fiducial beam down to a fractional RMS of $10^{-5}$.

To generate our perturbed beam model (i.e. the a priori unknown ``truth'' beam that we will aim to solve for from the data), we take the $\ell m$ cuts described above and add random Gaussian noise to them, tuned to create fluctuations in the beam amplitude at roughly the -30 dB level \citep{Fagnoni2021}.
\autoref{fig:beam_decomp} shows this perturbed beam, demonstrating the complex angular and frequency structure one might expect from a real antenna response located in the field. Furthermore, we show the frequency response of the beam, demonstrating the perturbed beam's more complex frequency structure relative to the fiducial model that looks visually representative of what has been seen in electromagnetic simulations \citep{Fagnoni2021}.
In total, the primary beam model holds 5 frequency degrees of freedom and 77 angular degrees of freedom (not including $\ell=0$) for a total of 385 parameters.

Due to the differentiable nature of the forward model, we can enact priors on the beam in both harmonic space on the $a_{\ell m}$ modes, as well as in real space where our intuition of the beam is actually gleaned.
In this work we set a Gaussian prior on the beam in real space centered at the fiducial beam with a variance that is tuned to yield fluctuations in the beam at the -25 dB level, demonstrated through prior predictive checks.
This is a fairly realistic assumption for real low-frequency telescopes \citep{Line2018, Nunhokee2020}, and basically says that while we may have confidence in our theoretical models of the primary beam near zenith, our knowledge of the far sidelobes is effectively unconstrained.

\subsection{Noise Model}
\label{sec:noise}

Thermal noise in the raw data is sourced at the visibility level, and is drawn from a complex-valued normal distribution that is assumed to be uncorrelated between different time bins, frequency bins, and baselines.
For thermal noise sourced at the amplifiers in the front end of a radio receiver, this is a very good approximation.
We assume a single noise variance for all times, frequenices, and baselines, with an amplitude that is tuned to yield a $\sim10\sigma$ power spectrum detection of our simulated EoR signal at $k\sim0.2\ h\ {\rm Mpc}^{-1}$.
This is representative of what an early detection by HERA might look like with a single observing season of data \citep{DeBoer2017}.
In other words, the covariance of the noise vector $\bm{n}$, which has the same dimensionality of $\bm{y}$ in \autoref{eq:rime_matrix}, has a covariance
\begin{align}
    \bm{N} = \langle \bm{n}\bm{n}^\dagger\rangle
\end{align}
that is diagonal and scalar, such that $N_{ii} = \sigma_n^2$.

A slightly more realistic noise model would entail simulating a total-power observation of the diffuse foreground sky to compute the measured sky temperature as a function of frequency and observing time, and adding this with a receiver temperature describing thermal noise originating from the front-end analog system \citep[as in][]{Aguirre2022}.
However, the simpler model adopted here allows us to fine tune the noise amplitude for diagnostic purposes, and is more than sufficient to demonstrate the proof-of-concept signal recovery studied in this work.

Note that the delay filtering step applied to the raw and model visibilities (\autoref{eq:rime_filter}) will slightly change the noise properties of the data, with an updated covariance given as
\begin{align}
\label{eq:filtered_noise}
\tilde{\bm{N}} = \bm{F}\bm{N}\bm{F}^\dagger.
\end{align}
While $\bm{N}$ was a diagonal matrix, $\tilde{\bm{N}}$ need not be, however, due to the fact that $\bm{F}$ is a very narrow high-pass filter, visual inspection shows $\tilde{\bm{N}}$ to be strongly diagonally dominant, and thus we maintain the usage of a diagonal covariance matrix but replace $N_{ii}$ with $\tilde{N}_{ii}$ in the likelihood (\autoref{eq:likelihood}).

To test the appropriateness of this approximation, we run a simple Monte Carlo test by generating random noise draws, filtering them with the specified delay filter, and then computing a histogram of the chi-square distribution, $\bm{\chi}^2=\bm{n}^\dagger\tilde{\bm{N}}^{-1}\bm{n}$.
We do this both for a full dense noise matrix (\autoref{eq:filtered_noise}) and a diagonal approximation of it.
The shape of the histograms are very similar, with shifts in their respective medians confined to less than 20\% of their estimated standard deviations. While this tells us that the approximation likely has a small or negligible impact on our inference, future work could explore this impact in more detail, for example by re-running inference with and without a dense noise covariance.

\section{Signal Estimation and Posterior Sampling}
\label{sec:signal_recovery}

In this section we demonstrate a proof-of-concept optimization and posterior sampling exercise given our mock HERA observation from a set of a priori unknown ``truth'' set of model parameters.
The goal is to optimize the joint model to the maximum a posteriori (MAP) value, and then to sample the posterior via a Markov Chain Monte Carlo (MCMC) process.
After sampling the posterior we are left with an approximation of it that will allow us to effectively marginalize the posterior across our foreground and instrumental nuisance parameters, thus acheiving our goal of characterizing the joint posterior distribution and performing end-to-end uncertainty propagation.

\subsection{Posterior Optimization}
\label{sec:optimization}

One of the challenges in optimizing a model like the one described above is the intrinsic degeneracy between different components of our data model.
For example, for the same frequency mode, the EoR and diffuse sky models are perfectly degenerate within the main field-of-view.
In practice, low-order frequency modes of the foreground model modulated by high-frequency modes of the beam model can also be partially degenerate with higher frequency modes of the EoR model.
This is not unique to our end-to-end forward model approach, and is indeed indicative of the challenge of the 21\,cm inverse problem.
These degeneracies create narrow valleys in the posterior that are difficult for the optimizer to navigate, especially in high dimensions.
In our experimentation, we have therefore not suprisingly found the most success by employing 2nd-order optimization routines like the L-BFGS quasi-Newton method over 1st-order approaches liks stochastic gradient descent.
The L-BFGS algorithm uses a sparse-Hessian approximation that allows it to better navigate ill-conditioned and high-dimensional parameter spaces like the one presented in this work \citep{Liu1989, Nocedal}.

\begin{figure}
\includegraphics[width=\linewidth]{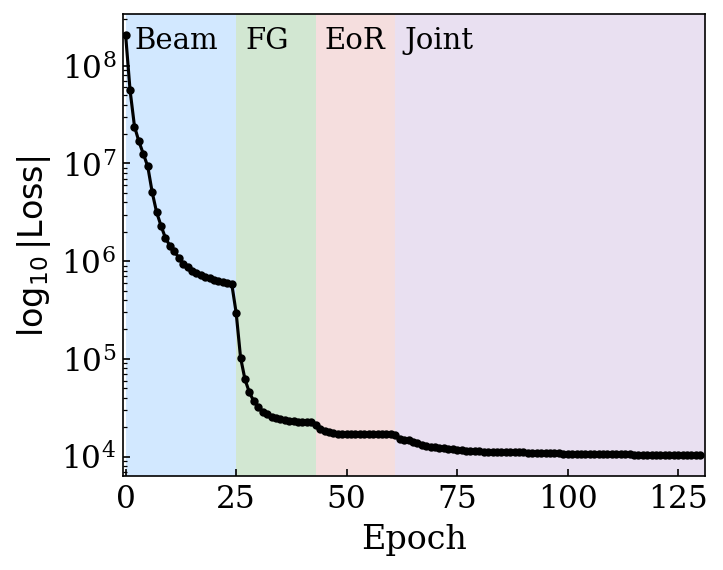}
\caption{Optimization of the forward model using the quasi-Newton L-BFGS solver, starting with disjoint optimization of each component, followed by a joint optimization. We have thinned the number of epochs for visual clarity. In total we run roughly 100 iterations for each component separately and then run roughly 1000 iterations jointly.}
\label{fig:optimization}
\end{figure}

In particular, we have found that there is a strong degeneracy between the $m=0$ mode of the 21\,cm sky model and the beam model.
Perhaps not surprisingly, this is due to the drift-scan nature of the simulated observations, where the $m=0$ mode of the sky acts as a constant offset in the visibilities as a function of observing time, which is degenerate with the combination of the beam and the foreground model.
As a consequence, we remove the $m=0$ modes of the 21\,cm model out of the optimization procedure because, without aggressive regularization, they can make the Hessian matrix singular.
This does not impact our ability to make an unbiased recovery of the EoR power spectrum, as the final power spectrum (described in \autoref{sec:ssfb_pspec}) is simply an average over $\ell\ \&\ m$ spherical harmonic modes, and we've effectively just set the $m=0$ weight to zero.

To further aid the convergence of the optimization, we first optimize each component independently before performing a joint optimization, running 100 iterations for each component before running roughly 1000 iterations with a joint parameterization.
We plot the results of the optimization in \autoref{fig:optimization}, showing the decrease in the loss function (in our case the un-normalized negative log posterior) as function of iterations.
We see that the beam optimization does the most to bring the model data and raw data into alignment, which highlights its importance in end-to-end signal estimation.
We terminate the optimization manually when the $k\sim0.1\ {\rm Mpc}^{-1}$ modes of the forward modeled EoR visibilities have stabilized.
After termination, the residuals formed by subtracting the model visibility data from the raw data are noise like, with no obvious spectral or temporal features.

Running the forward model in a data parallel manner spread across four NVIDIA A100 (80 GB VRAM) GPUs results in a runtime of $\sim0.3$ seconds for a single parameter update step, which involves a forward pass of the model and the backpropagation step.
Thus the total time for the optimization described above takes only a few minutes.

\begin{figure}
\includegraphics[width=\linewidth]{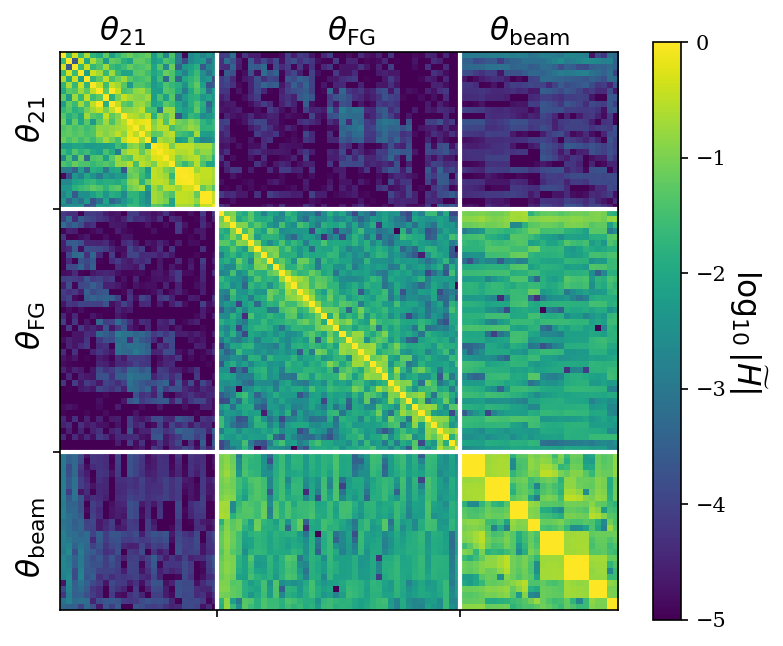}
\caption{A subset of the diagonal-normalized Hessian matrix across the three components of the forward model, $\widetilde{\bm{H}}$.
There are strong off-diagonal entries between the beam and foreground component, and weaker but still non-zero off-diagional entries between the beam and the 21\,cm component. Note this does not necessarily represent the cross-covariance between the components, which would be found by inverting the Hessian, but this does give intuition for the degeneracies between the components. Also note that this is only a small subset of the otherwise 80-thousand by 80-thousand Hessian matrix.
}
\label{fig:hessian}
\end{figure}

\begin{figure*}
\centering
\includegraphics[width=\linewidth]{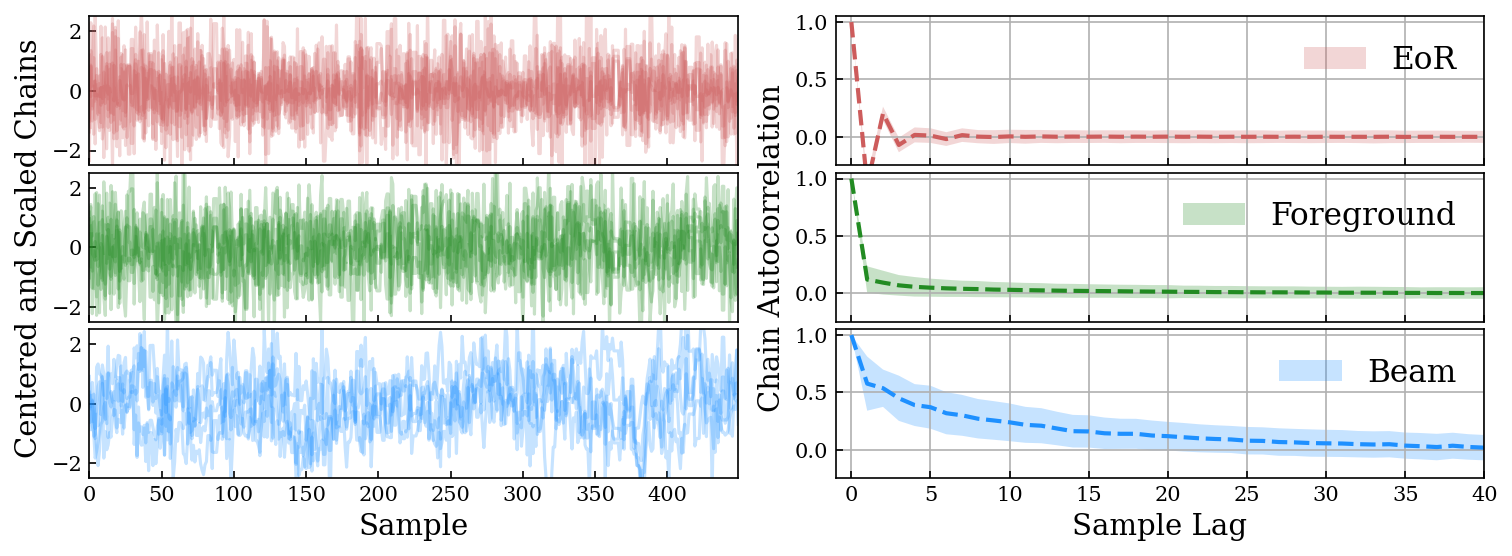}
\caption{Left: HMC-NUTS chains from \autoref{sec:hmc} for the three components of our data model (EoR, foreground, beam). Each chain that is shown represents a random parameter from its associated component, and is centered and scaled for visual clarity. Right: The average autocorrelation for all parameters of a given component (dashed) and their $\pm1\sigma$ region (shaded). Using \autoref{eq:ess} on the average autocorrelation for each component (dashed), we compute autocorrelation lengths of roughly (2, 3, 15) for the (EoR, foreground, beam) components, respectively.}
\label{fig:hmc_autocorr}
\end{figure*}

\subsection{Posterior Sampling}
\label{sec:sampling}

Once we've optimized to the maximum a posteriori (MAP) estimate, we'd like to quantify the shape and width of the posterior in order to perform uncertainty quantification.
One approximate way we can do this is by quadratically Taylor expanding the posterior about its MAP estimate using the Hessian matrix, which forms a Gaussian approximation to the posterior known as the Laplace approximation.
However, a Gaussian approximation to the posterior may be insufficient for noisy data or a posterior distribution that is multi-modal or has complex degeneracies.
More standard in the Bayesian inference literature is to sample the posterior via a Markov Chain Monte Carlo (MCMC) method.
In particular, the Hamiltonian Monte Carlo (HMC) approach \citep{Duane1987, Neal2011} and its variants such as the No-U-Turn sampler \citep[NUTS;][]{Hoffman2011} are considered state-of-the-art for complex, high-dimensional Bayesian inference problems.
These samplers simulate Hamiltonian dynamics in a dual position and momentum space to make Markov proposals that have low autocorrelation, and thus converge to the underlying posterior distribution more quickly than a random walk Metropolis-Hastings algorithm.
See also \citet{Jasche2013, Hernandez2021} for instances of HMC applied to cosmological parameter inference.
We refer the reader to \citet{Betancourt2017} for a review of HMC and NUTS.

Although HMC samplers are considered state-of-the-art for many Bayesian inference tasks \citep{Betancourt2017}, they still often struggle when tackling very high-dimensional and degenerate parameterizations found in real-world applications.
To confront these inference problems, it is beneficial to precondition the system with the posterior Hessian matrix, $\bm{H}$ \citep{Girolami2011}.
In the HMC literature, this is known as the Hamiltonian \emph{mass matrix}, $\bm{M}$, which defines the mapping between the momentum vector and the gradient of the position vector \citep{Neal2011}.
AD-enabled forward models are convenient in that they allow for explicit computation of the Hessian matrix using the computational graph itself.
However, even with automatically differentiable gradient calculations, it can still be difficult to compute, store, and invert the full Hessian matrix of the system.
As a consquence, it is common to see a diagonal mass matrix used to partially precondition the system.

For our case study, we have found that a diagonal approximations is not effective at enabling efficient exploration of the posterior distribution.
Therefore, we use a block-diagonal Hessian matrix to precondition the HMC sampling.
We compute dense Hessian matrices for each model component (e.g. EoR, FG, beam) while ignoring their off-diagonal terms, with the only exception being the $\theta_{\rm FG}\times\theta_{\rm beam}$ off-diagonal, which we keep because it is small in size and has outsized influence in the Hessian matrix.
We show a small subset of the diagonally normalized Hessian matrix in \autoref{fig:hessian} to demonstrate this.
This quantity, $\widetilde{\bm{H}} = \bm{h}^{-1/2}\bm{H}\bm{h}^{-1/2}$ where $\bm{H}$ is the Hessian matrix and $\bm{h}$ is the diagonal of the Hessian matrix, effectively normalizes the diagonal to be one, and thus makes it easier to visualize the importance of the off-diagonal components.
The Hessian matrix, being the matrix of second-order derivatives of the negative log posterior, shows strong off-diagonals between the beam and foregrounds, and weaker off-diagonals between the beam and the EoR.
Note this does not represent the cross-covariance between the different components of the forward model, which could be computed by inverting the Hessian matrix, but rather gives a sense for the degeneracies between the parameters.
Also note that this is only a small subset of the nearly 80 thousand parameters in the full Hessian matrix, and only goes to roughly show the importance of the block diagonal and off diagonals.

Note that to run HMC we only need the Cholesky factor of the adopted mass matrix.
In \autoref{sec:hmc} we review the quantities needed to simulate HMC trajectories and discuss how to do this in $\mathcal{O}(N^2)$ time given only the mass matrix Cholesky factor.
Future work will explore how to leverage redundant structures in the Hessian matrix to create sparser preconditioners that will enable scaling to even larger parameter dimensionalities.
\citet{Lentati2013, Sims2019} also propose a Hessian-preconditioned HMC sampler using an eigendecomposition of the Hessian matrix, which has a higher upfront cost than a Cholesky decomposition but can more easily handle singular matrices.

Having chosen our HMC-NUTS mass matrix, the final two parameters for HMC are the step-size and path-length.
We manually tune the step-size to be the largest possible while still returning high acceptance probability (greater than 90\%).
The path-length parameter is automatically resolved by NUTS' termination criterion \citep{Hoffman2011}.
In practice, we find that the HMC trajectories often terminate between 128 -- 256 steps.
Finally, we use the biased progressive sampling approach to sample the final ending point of the HMC trajectroy \citep{Betancourt2017}, which gives preference to points in the trajectory farther away from the initial point.

We run the sampler for 500 iterations, discarding the first 50 due to burn-in. 
In total the sampling process takes 16 hours to run across 4 GPUs, totalling to 64 GPU hours.
A visualization of the resultant HMC chains and their autocorrelation can be found in \autoref{fig:hmc_autocorr}, showing relatively low autocorrelation with an effective sample size (ESS) of over 100 for the EoR component.
The foreground component also maintains a low autocorrelation length, while the beam component sees a higher autocorrelation and thus a lower effective sample size (see \autoref{sec:hmc}).
We speculate that the longer autocorrelation length for the beam model is due to the non-linear absolute value operation applied to the beams during the forward modeling process, making the parameter space slightly more difficult to navigate.

\begin{figure}
\centering
\includegraphics[width=\linewidth]{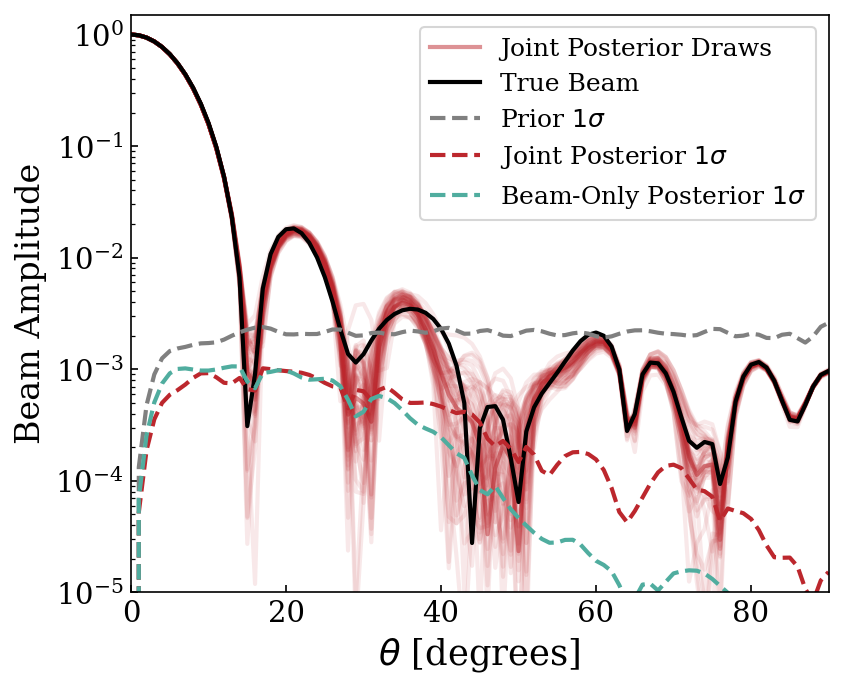}
\caption{The estimated posterior distribution of the beam at 125 MHz, cut along an azimuthal slice. We plot the true beam response (black) against a representative set of MCMC draws from the forward modeled beam posterior distribution (red solid), showing good agreement of the posterior draws with the underlying ground truth. We also plot the standard deviation of the forward modeled prior distribution (also known as the prior predictive distribution) of the beam (gray dashed), the jointly marginalized posterior predictive distribution (dashed red), and the beam-only marginalized posterior predicive distribution (blue dashed). We see that the posterior is tighter than the prior, as one would expect, and that the variance shrinks near the observer's horizon (for $\theta\rightarrow 90^\circ$), as we intuited before based on the effects of the high-pass delay filter. Furthermore, we see that the beam-only marginalized posterior is indeed tighter than the full, jointly marginalized posterior, indicative of non-neglible correlations between the beam and other components in the data model, as we suspected.}
\label{fig:beam_posterior}
\end{figure}

Visualizing an 80-thousand-dimensional posterior distribution is inherently difficult.
Instead of looking at the posterior for each parameter, we can forward model the posterior chains into a more intuitive intermediary representation, one that may allow for easier comparison with our prior knowledge.
For the beam, this would be its real-space representation, and for the EoR signal this could be its pixelized map representation, for example.
In \autoref{fig:beam_posterior}, we show draws from the posterior chains of the beam response, which represents the marginal posterior on the beam model.
We see that the posterior draws do a good job representing the true underlying beam response while also capturing the differing levels of uncertainty as a function of polar angle.
Regions where this agreement is worse, e.g. around $40^\circ$, are possibly due to lower sensitivity to the beam response at those angles.
Represented as dashed lines, we also compare the standard deviation of the beam's marginal posterior, the starting prior, and the conditional posterior (i.e. the posterior holding the foregrounds and EoR parameters constant).
As expected, we see the marginal posterior is tighter than the prior, but not as tight as the conditional posterior.
This tells us that we are indeed capturing the extra uncertainty in our beam model due to degeneracies between the beam and other components in our model, which act to inflate the variance on the beam parameters.

\begin{figure}
\centering
\includegraphics[width=\linewidth]{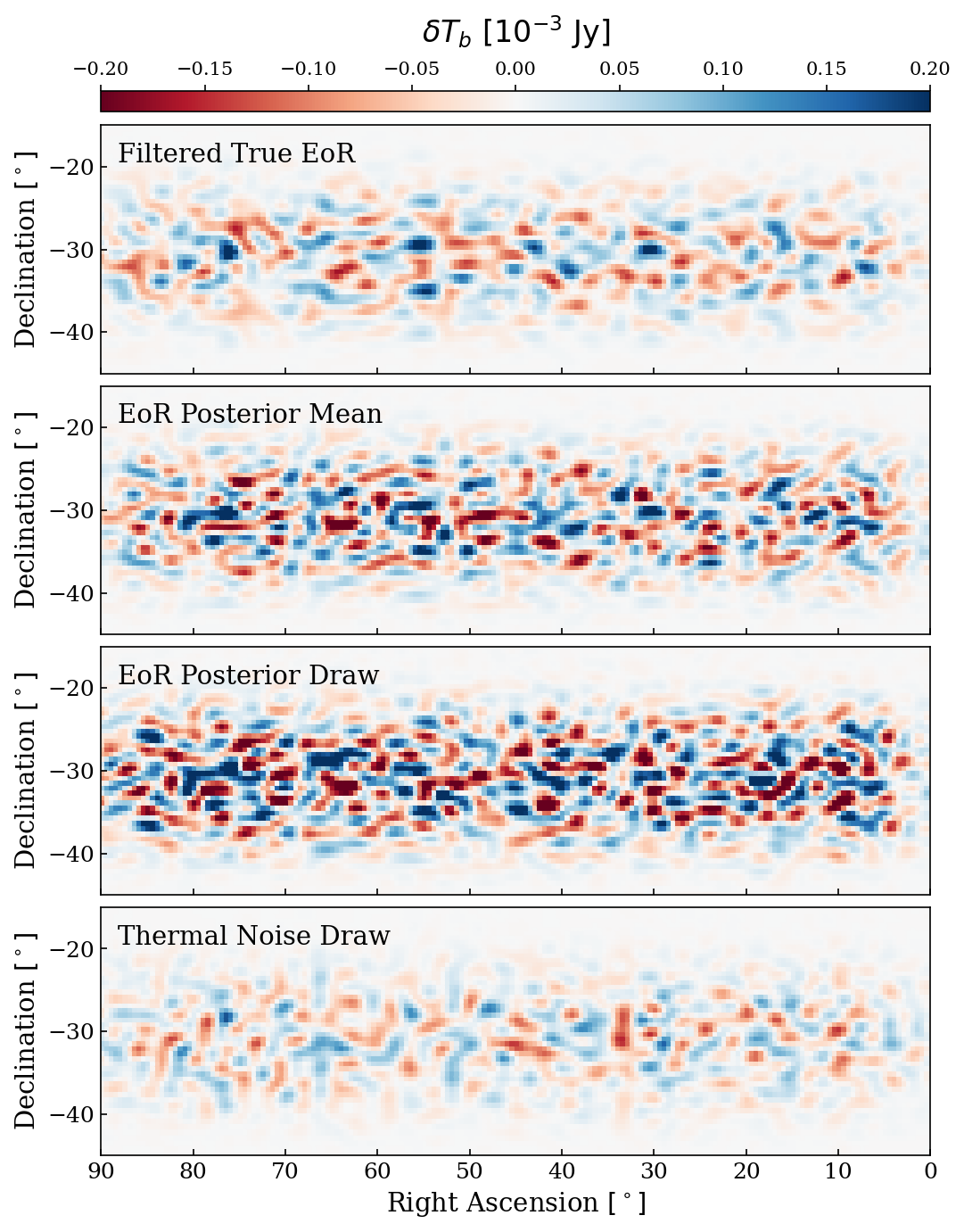}
\caption{Forward-modeled EoR maps at 125 MHz.
These maps come from forward modeling a signal to the full time-ordered visibilities and then applying the imaging step \autoref{eq:vis_to_map}.
The maps have not been primary beam corrected, so the preceived flux is attenuated near the image boundaries.
We show the true underlying EoR signal after high-pass visibility filtering (top), as well as the EoR map corresponding to the mean of the HMC posterior chains (middle-top) and a random draw from the HMC chain (middle-bottom).
We also image a thermal noise draw from our visibility noise covariance (bottom).
Relative to the true EoR map, the recovered posterior mean appears noisier due to a combination of the thermal noise in the maps as well as degeneracies with the foreground and instrumental components of the forward model.
}
\label{fig:eor_map_posterior}
\end{figure}

\begin{figure}
\centering
\includegraphics[width=\linewidth]{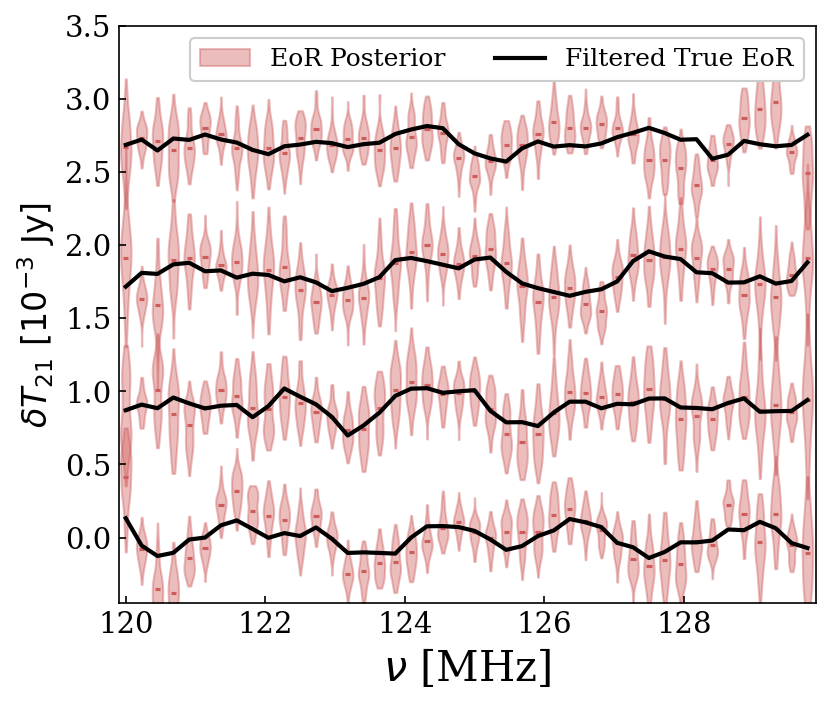}
\caption{The EoR component of the HMC posterior chains forward modeled and imaged into sky maps, showing a few sight-lines within the main field-of-view. We show the marginal EoR posterior distribution (red) alongside the underlying true EoR signal after delay-filtering and imaging (black). The width of the posterior is driven mainly by marginalized uncertainty from foregrounds and instrument parameters, with a subdominant component coming from the thermal noise in the raw data. The sight-lines have been artificially shifted for visual clarity. This demonstrates that we can estimate the 21\,cm posterior at the field level, which can then be projected to various summary statistics if desired.
We also test the statistical consistency of the posterior distributions with the true signal via a p-value test (described in the text), showing no evidence for significant statistical deviation of the filtered true EoR signal from the posterior chains.
}
\label{fig:eor_violins}
\end{figure}

Next we can inspect the posterior distribution marginalized over the beam and foreground components onto the EoR signal.
This is in effect probing the posterior of the 3D EoR signal at the field level, which allows us to capture uncertainty on the maps as well as any summary statistic we might care to form on top of these maps.
Recall that so far we have yet to define any kind of formal summary statistic for the EoR field: our optimization and sampling have simply leveraged the forward model that maps signals directly to the complex visibilities.
To visualize the EoR component of the MCMC chains we take each EoR sample in the chain and forward model it to the visibility level.
Then we apply an imaging step to turn it into a wide-field map.
We discuss the mathematics of this step in \autoref{sec:power_spectra} but we will discuss the results here.

\autoref{fig:eor_map_posterior} shows an example of the maps produced by this process at 125 MHz.
Note that the maps have not been primary beam corrected, so they will be naturally attenuated by the edges of the image.
We show the true underlying EoR signal imaged after applying the delay filter (top), along with the marginal posterior mean (middle-top), a random draw from the posterior (middle-bottom), and a realization of the thermal noise in the data (bottom).
We see there is more effective noise in the posterior mean and draws that comes from the marginalization of uncertainty from the foreground and instrumental parameters.
Looking carefully, one can see that the rough features of the filtered true EoR map are indeed preserved in the posterior mean image, particularly nearly the maximum response of the telescope at a declination of -30.72 degrees.

Because we are producing 3D images of the EoR sky signal, we can also look at the data along a line-of-sight at a fixed right ascension and declination.
Recall that for an intensity mapping probe the line-of-sight direction is directly mapped to the observing frequency of the telescope.
\autoref{fig:eor_violins} shows a few random sightlines near the peak response of the telescope.
We plot the true underlying EoR signal after applying the delay filter (black) alongside the full marginalized posterior (red).
This better represents the fact that we are indeed probing the posterior of the EoR signal at the field level, whose per-frequency averages show high correlation with the underlying signal in the data.
We see that the posterior means (red dashes) are in good agreement with the underlying true EoR signal, demonstrating that we can accurately recover both the amplitude and phase of the EoR maps at the field level.

We can also describe the agreement between the true, filtered EoR signal in the data and the recovered posterior chains a little more quantitatively.
To do this, we take the EoR map posterior draws and compute an empirical frequency-frequency covariance matrix, $\Sigma$, averaging over all posterior draws and over all pixels near the maximum sensitivity of the beam at a declination of $\sim-30.7^\circ$ (roughly 250 pixels).
We then draw 1000 mean-zero realizations from this covariance and compute the chi-square statistic, $\chi^2=r^T\Sigma^{-1}r$, for each one, where $r$ is a single draw.
This gives us the empirical $\chi^2$ sampling distribution of our EoR map sight-lines within the main field-of-view.
We then take the four sight-lines in \autoref{fig:eor_map_posterior}, subtract from them the posterior mean sight-line at each pixel value, and compute the same chi-square statistic.
We can then posit a null hypothesis that the true EoR sight-lines (black) can be descrbed as a random draw from the empirical covariance derived from the posterior chains.
The probability this null hypothesis is rejected, or the p-value, is then computed as the fraction of our empirical chi-square sampling distribution that falls above the chi-square of each sight-line.
For the four sight-lines plotted in \autoref{fig:eor_violins} these take values of [0.34, 0.26, 0.72, 0.27], which would not suggest a rejection of our null hypothesis.
Going a step further and computing this for all $\sim250$ sight-lines reveals a p-value probability distribution that is effectively uniform from 0 to 1, which is what one would expect for a signal drawn from the empirical covariance model.

\subsection{Map-making and Power Spectrum Estimation}
\label{sec:power_spectra}

We use the power spectrum as a summary statistic, which is both well understood and holds a significant amount of the information content in the Cosmic Dawn 21\,cm signal \citep{Prelogovic2024}, although future work could explore alternative summaries that exploit non-Gaussian information.
There are multiple ways to estimate the 21\,cm power spectrum given a set of interferometric visibilities.
Some estimators, such as the delay spectrum discussed in \autoref{sec:fg_problem}, go straight from the visibiltiies to the power spectrum, while other approaches first reconstruct the sky via a map-making process and then estimate the power spectrum from those maps.
There is an expansive literature on interferometric map-making for 21\,cm cosmology \citep[e.g.][]{Sullivan2012, Shaw2014, Dillon2015a, Eastwood2018, Morales2019, Xu2024}, which we will not review in depth here and instead refer the reader to \citep{Liu2020} for detailed discussions.
Recall that all of the optimization and posterior sampling described above relies only on the forward pass of the model (from sky to visibilities) and the backpropagation algorithm.
Having already performed the optimization and sampling, we will now use images and power spectra to quantify the results.

\subsubsection{Map-making}
Briefly, the map-making step produces images of the sky by transposing the forward transformation of the visibility simulation (\autoref{eq:rime_matrix}) and multiplying by a user-defined normalization matrix.
Let the noisy visibility ($\bm{y}$) for a single frequency channel be written as
\begin{align}
\label{eq:map_to_vis}
\bm{y} = \bm{A}\bm{x} + \bm{n},
\end{align}
where $\bm{n}$ is Gaussian noise.
Then the generalized map-making solution is defined as
\begin{align}
\label{eq:vis_to_map}
\hat{\bm{x}} = \bm{D}\bm{A}^\dagger\bm{N}^{-1}\bm{y},
\end{align}
where $\hat{\bm{x}}$ is the estimated map, $\bm{A}$ is the matrix encoding the beam and fringe response in \autoref{eq:rime_matrix}, and $\bm{D}$ is a user-defined invertible normalization matrix \citep{Tegmark1997, Dillon2015a}.
The noise in the visibilities is assumed to be drawn from a mean-zero, uncorrelated Gaussian distribution with a covariance of $\bm{N} = \langle \bm{n} \bm{n}^\dagger\rangle$.
The choice of normalization matrix $\bm{D}$ depends on the desired statistical properties of the map.
We can further write down the point spread function (PSF) of the maps,
\begin{align}
\label{eq:map_psf}
\bm{P} = \bm{D}\bm{A}^\dagger\bm{N}^{-1}\bm{A},
\end{align}
which, under an ensemble average of the maps $\langle \hat{\bm{x}}\rangle$,  satisfies the following relation
\begin{align}
\langle \hat{\bm{x}} \rangle = \bm{P}\bm{x}.
\end{align}
Thus $\bm{P}$ describes how the measurement process of the interferometer mixes the intrinsic flux of a map pixel with neighboring pixels.
The ``optimal'' choice of $\bm{D}$ depends on the desired statistical properties of $\hat{\bm{x}}$, but generally the optimal map-making formalism refers to a collection of approaches that retain all of the statistical information encoded in $\bm{y}$.
In theory one would choose the maximum likelihood solution $\bm{D}=(\bm{A}^\dagger\bm{N}^{-1}\bm{A})^{-1}$, but this is almost never strictly invertible for radio interferometers and thus a range of alternatives exist.
Note that if we wanted to include the high-pass Fourier filtering of the visibilities described in \autoref{sec:rime} then the PSF matrix becomes $\bm{P} = \bm{D}\bm{A}^\dagger\tilde{\bm{N}}^{-1}\bm{F}\bm{A}$, where recall $\bm{F}$ is the filtering operation, and now $\bm{D}$, $\bm{A}$, $\tilde{\bm{N}}$, and $\bm{y}$ are stacks of themselves for each frequency bin.
Many authors choose a simple diagonal normalization matrix that, although does not deconvolve the map, is computationally efficient, and so long as we can compute $\bm{P}$ we can always make faithful comparisons to models of the sky \citep{Dillon2015a}.
In this work we also use such a diagonal normalization matrix

\subsubsection{Power Spectrum Estimation}

Having produced maps of the sky we are now prepared to compute their power spectra.
Under a flat sky approximation we could take the 2D transverse Fourier transform to generate $\bm{k}_\perp$ modes and a 1D line-of-sight transform to generate $\bm{k}_\para$ modes, however, for large fields-of-view this relationship breaks down as a single line-of-sight does not exist.
The appropriate generalization of 3D Fourier transforms on the sphere is the spherical Fourier-Bessel (SFB) formalism, used extensively in wide-field galaxy survey analyses \citep[e.g.][]{Binney1991,Leistedt2011,Rassat2012, Pratten2013,Gebhardt2021}, and recently adapted for intensity mapping experiments \citep{Liu2016c}.
Here we will use the SFB formalism for power spectra estimation, but do so under the newly defined spherical stripe Fourier-Bessel (SSFB) formalism, which we introduce in \autoref{sec:ssfb}.

\begin{figure*}
\centering
\includegraphics[width=\linewidth]{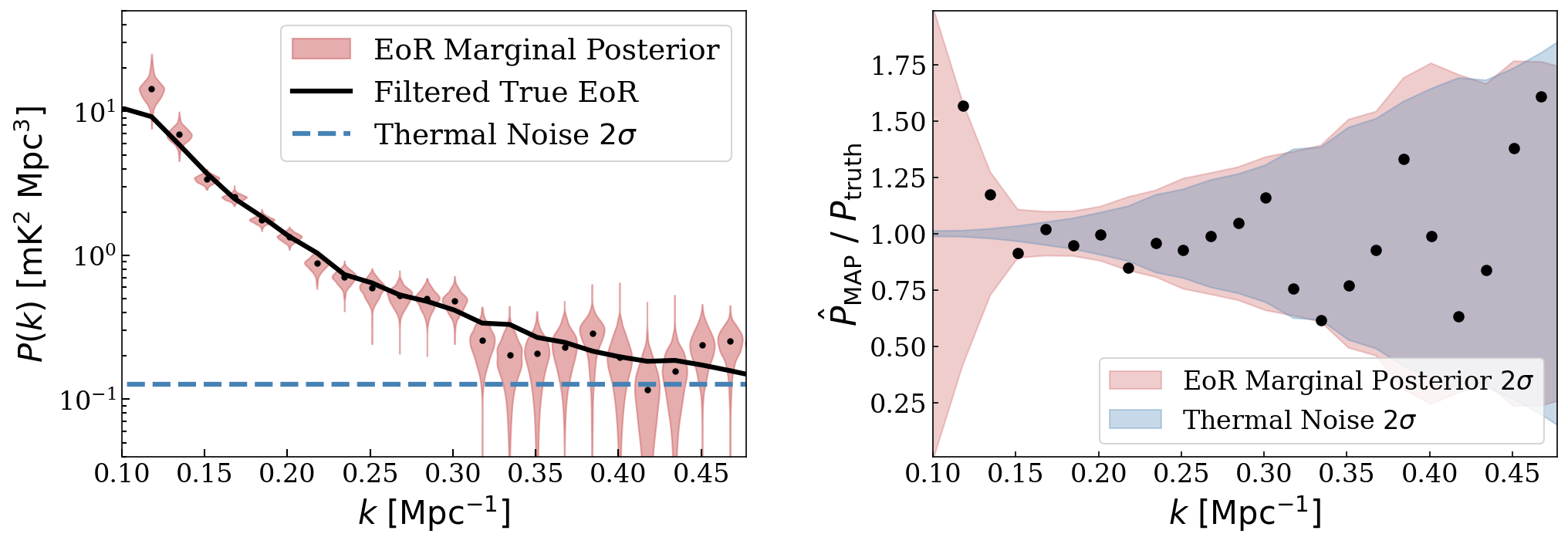}
\caption{
The EoR power spectrum posterior.
{\bfseries Left}: Power spectra from forward modeling draws from the HMC chain (red), representing the marginal posterior distribution on the EoR power spectrum. These distributions are in good-agreement with the underlying true EoR power spectrum down to the delay-filtering scale of $k=0.1\ {\rm Mpc}^{-1}$, below which the data are attenuated by the filter. We also show the averaged $2\sigma$ level from our mock noise draws, which represents the noise floor in the recovered power spectra (blue-dashed).
{\bfseries Right}: The maximum a posterior EoR power spectrum divided by the true filtered EoR power spectrum (points). We also show the $\pm2\sigma$ noise distribution (blue) and EoR marginal posterior (red), the latter of which demonstrates good agreement with the fractional errors observed in the recovered power spectra across all $k$ modes up to $k\sim0.5\ {\rm Mpc}^{-1}$, above which the data are noise dominated.
The sharp increase in uncertainty for $k<0.15\ {\rm Mpc}^{-1}$ is due to the marginalization of uncertainty from the foreground and instrument model onto these modes, achieving our stated goal of computing a fully ``end-to-end'' errorbar on the power spectrum.
}
\label{fig:eor_pspec_posterior}
\end{figure*}

In \autoref{sec:ssfb_pspec} we specifically discuss power spectrum estimation within the SSFB formalism, which we will briefly review here.
Let the 21\,cm temperature field be $T(\bm{r})$ in units of Kelvin.\footnote{We typically express the sky brightness distribution in units of specific intensity, or Jansky/steradian, but at radio frequencies we can also equivalently express it as a temperature using the Rayleigh-Jeans law.}
The 3D Fourier transform of the field is written as
\begin{align}
\label{eq:T_FT}
\widetilde{T}(\bm{k}) = \int d^3r\ e^{i\bm{k}\bm{r}}\ T(\bm{r}),
\end{align}
with the inverse transform $T(\bm{r}) = \mathcal{FT}^{-1}(\widetilde{T}(\bm{k}))$ picking up units of $1/(2\pi)^3$ for normalization.
The power spectrum is defined as the square of the Fourier-transformed field under an ensemble average, given as
\begin{align}
\langle \widetilde{T}(\bm{k})\widetilde{T}^\ast(\bm{k}^\prime)\rangle = (2\pi)^3\delta^D(\bm{k}-\bm{k}^\prime)P(\bm{k}),
\end{align}
where $\delta^D(\bm{k}-\bm{k}^\prime)$ is the Dirac delta function. 
Thus we often think of the power spectrum as the square of the Fourier-transformed field.

In the spherical Fourier-Bessel formalism, we have a different representation of the field in Fourier space, one that is given as
\begin{align}
T_{\ell m}(k) = \int d\Omega dr\ r^2j_\ell(kr) Y_{\ell m}^\ast(\hat{r}) T(\hat{r}, r),
\end{align}
where $T_{\ell m}(k)$ are the SFB coefficients, $d\Omega={\rm sin}\theta d\theta d\phi$, $Y_{\ell m}$ are the spherical harmonics, and $j_\ell$ is the spherical Bessel function of the first kind.
To estimate $P(\bm{k})$ using the spherical Fourier-Bessel formalism, we need an analogous relationship between the SFB-transformed field and the power spectrum, which is given in \citep{Liu2016c} as
\begin{align}
\label{eq:sfb_pspec1}
\langle T_{\ell m}(k)T^\ast_{\ell^\prime m^\prime}(k^\prime)\rangle = k^{-2}\delta^D(k-k^\prime)\delta_{\ell\ell^\prime}\delta_{m m^\prime}P(k).
\end{align}
Similar to before, we see that we can estimate the power spectrum directly by squaring and binning SFB modes estimated from the maps. 
In practice, to do this numerically we create SFB transformation matrices that map the pixelized sky to the SFB coefficients and then square them and average them to estimate the 1D power spectrum $P_k$.
We use a Hann function to apodize the maps along the frequency direction before taking the SSFB transform, which reduces Fourier-space sidelobes when taking the SFB transformation.

When forming the power spectrum as in \autoref{eq:sfb_pspec1}, if the two temperature fields $T_{\ell m}(k)$ and $T_{\ell^\prime m^\prime}^\ast(k^\prime)$ are drawn with the same thermal noise then we will end up with an additive noise bias term in the power spectrum \citep{Dillon2014}.
This can be removed by cross-multiplying maps with different noise statistics, such that they average to zero, or it can be subtracted directly from the final power spectrum given an estimate of the bias.
In this work we use the latter approach, using a handful of simulated noise visibilities drawn from the noise covariance to estimate this noise bias term.
Note that we never include the \emph{actual} noise realization in this bias subtraction, only noise realizations drawn from the same covariance.

\subsubsection{The 21 cm Power Spectrum Posterior}
\label{sec:HI_pspec_posterior}

We are now prepared to take our HMC posterior chains on the EoR component and forward model them to the visibilities, image them into maps, and then estimate their SSFB power spectra.
In the left panel of \autoref{fig:eor_pspec_posterior} we show the derived marginal posterior on the power spectrum bins (red shaded) and their MAP estimates (black points) alongside the underlying filtered true EoR power spectrum (black line), which we truncate below $k\le0.1\ {\rm Mpc}^{-1}$ due to attenuation of the visibility delay filer (\autoref{fig:filtered_eor}).
For modes above the filter scale, we see good agreement between the recovered posterior distributions and the true power spectrum, which bottoms-out to the noise floor for $k\geq0.5\ {\rm Mpc}^{-1}$.

The agreement between the estimated power spectrum and the true underlying power spectrum is better captured in the right panel of \autoref{fig:eor_pspec_posterior}.
Here, we show the ratio of the maximum a posteriori (MAP) EoR power spectrum to the true EoR power spectrum (black points), with the shaded regions indicating the marginal posterior $2\sigma$ width (red) and the standard thermal noise $2\sigma$ width (blue).
As noted previously, most current 21\,cm analyses only compute the thermal noise uncertainty \citep{HERA2023}, which is relatively straightforward to compute, and are unable to account for an end-to-end uncertainty model that we now show here for the first time (red).
As expected, this uncertainty increases at a certain $k$ scale (in this case $k\sim0.15\ {\rm Mpc}^{-1}$), below which the EoR model becomes increasingly degenerate with the combined foreground and beam model.
To our knowledge, this is the first demonstration of an end-to-end, marginalized posterior distribution on the 21\,cm power spectrum accounting for fluctuations in both foreground parameters and instrumental parameters.

In a practical analysis, one would opt to use this joint uncertainty model when claiming a putative power spectrum detection, making it more robust to the threat of partially degenerate foreground and instrument parameters.
For example if we simply used the thermal noise uncertainty model for the power spectrum bins between $0.1 < k < 0.15\ {\rm Mpc}^{-1}$, the right panel of \autoref{fig:eor_pspec_posterior} shows we would produce biased measurements by upwards of $10\sigma$.
This is particularly important because these Fourier modes are also the modes that, for many theoretical EoR models \citep{Mesinger2011}, have the largest signal-to-noise ratio.
This means they make up the bulk of our total sensitivity and tend to drive astrophysical parameter inference \citep{Breitman2024}, making their robust estimation even more important.

\section{Conclusion}
\label{sec:conclusion}

Given the difficulty in separating the 21\,cm signal from bright foregrounds and complex instrumental systematics, recent years have seen a fresh wave of attention paid to more robust uncertainty quantification, systematic modeling, and the application of Bayesian methods \citep[e.g.][]{Sims2019, Rapetti2020, Anstey2021, Byrne2021, Burba2023, Kennedy2023, Scheutwinkel2023, Murphy2024, Pagano2024, Glasscock2024}.
The overarching aim of these approaches is to be able to more accurately and robustly subtract foregrounds and systematics while also accounting for their degeneracies with the underlying 21\,cm signal.
However, to date, a unified, end-to-end Bayesian forward model that can marginalize over both foregrounds and the telescope response for an interferometric experiment has yet to be proposed.
In this work we set out to develop the first end-to-end, differentiable Bayesian forward model for 21\,cm cosmology and line intensity mapping (LIM) more generally, called \texttt{BayesLIM}.
The framework aims to estimate the joint posterior distribution between a 3D cosmological signal alongside the often degenerate and poorly constrained foreground and instrumental response.
This is particularly important for 21\,cm LIM due to the presence of overwhelming foregrounds.
Although computationally demanding, we show that the advent of high-level automatic differentiation libraries combined with large-memory GPU computing can make the end-to-end Bayesian forward model a feasible solution in certain cases. 

We provide a proof-of-concept on a mock HERA observation where we model the antenna primary beam's frequency and angle-dependent response alongside a full-sky foreground sky model and a 21\,cm sky model.
In total the data model contains roughly 80,000 complex-valued parameters spanning the three components, with priors that are representative of our current best-understanding of the foreground sky and of low-frequency antenna primary beams.
We show we can optimize to an unbiased maximum a posteriori (MAP) solution for EoR Fourier modes $k > 0.1\ {\rm Mpc}^{-1}$, and also demonstrate end-to-end uncertainty quantification via Hamiltonian Monte Carlo sampling that for the first time can marginalize the uncertainty from both foreground and instrumental parameters onto the EoR Fourier modes.
The framework presented here will be key in moving the current state of 21\,cm cosmology from setting upper limits to making direct, robust detections of the 21\,cm cosmological signal, both for the 21\,cm power spectrum and the 21\,cm monopole.

In addition, we have presented a novel extension to the spherical harmonic formalism specific to sky masks used by drift-scan radio telescope observations, which we call the \emph{spherical stripe harmonics}.
These harmonics are a band-limited complete and orthogonal basis on the spherical stripe cut sky, and enable an order of magnitude reduction in the number of parameters needed to model a 21\,cm sky signal for HERA.
We also introduce their 3D analog, the spherical stripe Fourier Bessel formalism (SSFB) for modeling 3D intensity mapped signals and computing power spectra.

While the \texttt{BayesLIM} codebase is expanded and the framework applied to real experimental data, there are a number of areas for improvement going forward.
In particular, the inclusion of instrumental systematics like mutual coupling and gain calibration terms could be included for a more realistic end-to-end instrument model.
Furthermore, simulating the impact of missing data, caused by radio frequency interference, can be included and dealt with by inflating the variance of those pixels such that they have effectively no impact on the loss function.
Second, foreground models that include polarized sources as well as bright, extended radio sources like Fornax A will improve model realism when applied to real data.
In addition, more efficient posterior sampling, e.g. with neural posterior estimation \citep{Zeghal2022}, will help to tackle larger datasets such as a full HERA array or the SKA-core array.
Additionally, the inclusion of a differentiable astrophysical model, for example an emulated cosmological simulation, would enable direct constraints on astrophysical parameters and would make the priors on our cosmological signal basis more physically-motivated.
Finally, the power of an end-to-end differentiable framework like \texttt{BayesLIM} expands beyond just single-experiment 21\,cm signal inference.
To begin with, the model is perfectly amenable to combined 21\,cm monopole (i.e. global signal) and interferometric inference, where these could come from the same telescope or from different telescopes in different locations on Earth \citep[e.g.][]{Anstey2023}.
Futhermore, this can be extended to multi-tracer LIM analyses, where the shared model is the underlying cosmological density field and two separate signal and instrument models are created to jointly analyze 21\,cm data alongside another overlapping LIM tracer.

GPU acceleration is key to the feasibility of this approach on realistic interferometric datasets.
Assuming modest improvements to the efficiency of the forward model, based on current benchmarks (\autoref{sec:scaling}), it is reasonable to expect that a similar analysis presented here could be repeated on real HERA data (or future SKA-core data) with less than 1k (10k) GPU hours of compute, assuming similar baseline cuts but an increased number of frequency channels and time integrations.
These benchmarks put the performance of \texttt{BayesLIM} on par with other state-of-the-art GPU-based 21\,cm forward model simulators \citep[e.g.][]{Kittiwisit2025, Ohara2025}, with the added benefit of \texttt{BayesLIM}'s automatic differentiation backend.
It remains to be seen how much this computational budget will grow when additional components are added to the forward model, such as antenna calibration terms, mutual coupling systematics, or higher resolution sky models.
Nevertheless, there are a wide range of scientific LIM analyses, even ones short of a full end-to-end signal extraction analysis, that will benefit from more statistically rigorous, accelerated, differentiable Bayesian forward models like the one presented in this work.

\section*{Acknowledgements}

The author gratefully acknowledges the following individuals for supportive discussions spanning the duration of this work, including Benjamin Wandelt, Adrian Liu, Jacqueline Hewitt, Aaron Parsons, and Francois Lanusse.
The author gratefully acknowledges support from the MIT Pappalardo fellowship, as well as from NASA through the NASA Hubble Fellowship grant \#HST-HF2-51533.001-A awarded by the Space Telescope Science Institute, which is operated by the Association of Universities for Research in Astronomy, Incorporated, under NASA contract NAS5-26555.

\section*{Data Availability}

Data used in this work may be made available upon reasonable
request to the corresponding author.



\bibliographystyle{mnras}
\bibliography{main} 



\appendix

\section{The Spherical Stripe Fourier Bessel Formalism}
\label{sec:ssfb}

Here we describe the construction of an orthogonal set of three-dimensional modes in spherical coordinates defined over an observing mask suitable for drift-scan radio telescopes.
The novelty here is the derivation of a new set of angular modes on a spherical stripe of arbitrary polar extent and nearly arbitrary azimuthal extent, which we call the spherical stripe harmonics (SSH).
This formalism is inspired by the spherical cap harmonic (SCH) analysis developed for geophysics \citep{Haines1985, Thebault2006, Torta2019} and more recently suggested for use in 3D cosmological surveys \citep{Samushia2019}.
Like standard spherical harmonics operating over the full sky, spherical cap harmonics are orthogonal and complete over a cut sky confined to a polar cap on the sphere.
They were originally derived by \citet{Haines1985}, who solved the self-adjoint Sturm-Liouville problem for Laplace's equation defined over a polar cap, whose eigenfunctions are gauranteed to be band-limited complete and orthogonal.
The new spherical stripe harmonics presented here build upon this by treating both the minimum and maximum polar extent of the observing mask as free parameters.
The reason for pursuing a spherical harmonic basis tailored to a specific observing mask is mainly the need for parameter sparsity: fewer parameters means a smaller posterior dimensionality, which can yield significant improvements in computational efficiency when exploring the high dimensional parameter space.

Incorporating these new harmonics into the spherical Fourier Bessel formalism, which we now call the spherical stripe Fourier Bessel (SSFB) formalism, allows for more efficient representation of 3D cosmological fields.
The main drawback of this approach is the computational demand in computing the new harmonic modes on the sky: computing non-integer degree Legendre polynomials to high precision requires arbitrary precision computations that are generally slow, especially when attempting to generate all harmonic modes down to fine spatial scales.
However, when using these in a forward model that is to be evaluated many times with the same harmonic basis functions, one can pre-compute and cache the functions, which only incurs a one-time computational cost.

\subsection{Overview: The Spherical Fourier Bessel Formalism}

The spherical Fourier Bessel (SFB) decomposition is a means of representating a three dimensional field in harmonic space using basis vectors that are solutions to the Helmholtz differential equation. They have been used extensively for the representation of 3D cosmological data including galaxy surveys, weak lensing surveys, and more recently for intensity mapping surveys \citep{Binney1991, Heavens1995, Rassat2012, Liu2016c, Samushia2019, Gebhardt2021}. The advantage of the SFB approach is its natural ability to incorporate curved-sky effects from wide field-of-view surveys where the flat-sky approximation breaks down.

A generalization of Laplace's equation, the Helmholtz equation is written acting on a scalar field $f$ as
\begin{align}
\nabla^2 f + k^2 f = 0,
\end{align}
where $k$ is the wavevector of a wave propagating through the field.
The general solution to the Helmholtz equation in spherical coordinates can be broken into radial, polar, and azimuthal solutions, the latter two of which combine to form the well-known spherical harmonics.
This general solution can then be written in spherical coordinates as \citep[e.g.][]{Samushia2019}
\begin{align}
f_{lmk}(r,\theta,\phi) =\ &g_{lk}(r)Y_{lm}(\theta,\phi) = g_{lk}(r)\Theta_{lm}(\theta)\Phi_m(\phi)\nonumber \\ \nonumber
=\ &[A^j_{l}j_l(kr) + A^y_{l}y_l(kr)]\times\\ \nonumber
&[A^P_{lm}P_{l}^m(\cos\theta) + A^Q_{lm} Q_l^m(\cos\theta)]\times\\
&[A^+_me^{im\phi} + A^-_me^{-im\phi}],
\end{align}
where $Y_{lm}$ are the spherical harmonics, $j_l$ and $y_l$ are the spherical Bessel functions of the first and second kind, $P_l^m$ and $Q_l^m$ are the associated Legendre polynomials of the first and second kind, and the $A$s are a series of coefficients \citep[][Table 9.2]{Arfken}.
In the following, we will redefine the general solution by dividing by the first $A$ coefficient of the radial, azimuthal and polar terms to get
\begin{align}
f_{lmk}(r,\theta,\phi)
=\ &A_{l}^g[j_l(kr) + \tilde{A}^y_{l}y_l(kr)]\times \nonumber \\
&A_{lm}^Y[P_{l}^m(\cos\theta) + \tilde{A}^Q_{lm} Q_l^m(\cos\theta)]\times \nonumber \\
&[e^{im\phi} + \tilde{A}^-_me^{-im\phi}].
\end{align}
The consequence of this is to fold the overall normalization of the radial and angular solutions into $A_{l}^g$ and $A_{lm}^Y$, which we can compute after-the-fact given their respective orthonormality conditions.
We can then express the general spherical Fourier Bessel solution in compact notation as
\begin{align}
\label{eq:sfb_general}
f_{lmk}(\hat{r}, r) = g_{l}(kr)Y_{lm}(\hat{r}).
\end{align}
Note that although we denote $l$, $m$, and $k$ as subscripts we have yet to specify their values, and for the time being assume $m,k\in\mathbb{R}$ and $l\in\mathbb{R}\ |\ l\ge0$ until we specify their orthogonality and boundary conditions, at which point they will become discretized.

The relationship between real space (i.e. map space) and harmonic space (i.e. coefficient space) is dictated by the forward (harmonic to map space) and reverse (map to harmonic space) SFB transforms.
The reverse angular transform is known as the spherical harmonic transform (SHT)
\begin{align}
\label{eq:SHT}
T_{lm}(r) = \int d\Omega Y^\ast_{lm}(\hat{r})T(\hat{r}, r),
\end{align}
where $d\Omega=\sin\theta d\theta d\phi$ is the angular differential integrated across the full-sky.
The (orthonormalized) spherical harmonics satisfy the following orthogonality condition
\begin{align}
\label{eq:sht_ortho}
\int d\Omega Y_{lm}(\hat{r})Y_{l^\prime m^\prime}^\ast (\hat{r}) = \delta_{ll^\prime mm^\prime}.
\end{align}
The forward transformation from coefficient to map space is then written as
\begin{align}
T(\hat{r}, r) = \sum_{lm}Y_{lm}(\hat{r})T_{lm}(r),
\end{align}
where the sum is over $-l\le m\le l$ and $0 \le l < \infty$, athough in practice we truncate the sum over $l$ at some $l_{\rm max}$.

The reverse radial transform is known as the spherical Bessel transform (SBT)
\begin{align}
\label{eq:SBT}
T_{lm}(k) = \sqrt{\frac{2}{\pi}}\int_{r_1}^{r_2} dr\ r^2 k g_{l}(k r)T_{lm}(r).
\end{align}
The spherical Bessel radial modes satisfy the orthogonality condition
\begin{align}
\label{eq:SBT_ortho}
\int_{r_1}^{r_2} dr\ r^2 g_l(k r) g_l(k^\prime r) = \frac{\pi}{2}\frac{1}{kk^\prime}\delta_{kk^\prime}.
\end{align}
The forward radial transform therefore is
\begin{align}
T_{lm}(r) = \sqrt{\frac{2}{\pi}}\int dk\ k g_l(kr) T_{lm}(k).
\end{align}
Thus the full reverse and forward SFB transforms can be written as
\begin{align}
\label{eq:sfb_reverse}
T_{lm}(k) &= \sqrt{\frac{2}{\pi}}\int_{r_1}^{r_2}dr\ k r^2\int d\Omega\ f_{lmk}^\ast(\hat{r}, r) T(\hat{r}, r) \\
\label{eq:sfb_forward}
T(\hat{r}, r) &= \sqrt{\frac{2}{\pi}} \int dk\ k \sum_{lm}f_{lmk}(\hat{r}, r)T_{lm}(k),
\end{align}
where $f_{lmk}(\hat{r}, r)$ is defined in \autoref{eq:sfb_general}.
Computationally, the relative order of the SHT and SBT can be interchanged or done simultaneously, with implications for accuracy and computational speed of the SFB transform depending on the nature of the survey \citep{Leistedt2011}. 

Normally, $l$ and $m$ are constrainted to be integer-valued, set by the boundary condition that the Legendre polynomials are well-behaved at the poles ($\theta=0,\pi$), and the adoption of the azimuthal periodicity boundary condition.
This also allows for the polynomials to be written in recursive form as derivatives of the standard Legendre functions of integer degree.
However, the associated Legendre functions can be re-written in a more general form allowing for non-integer degree $l \ge 0$ and non-integer order $|m|\le l$ using the Gaussian hypergeometric function \citep{Cohl2020}.
In this form, both $P_l^m(x)$ and $Q_l^m(x)$, and their derivatives, are well-defined on the interval $|x|< 1$ for non-integer order and degree, and are also known as Ferrers function of the first and second kind. Indeed, this is the approach for generating the aforementioned spherical cap harmonics.
Note that while $P_l^m(x)$ contains a regular singularity at $|x|= 1$, $Q_l^m(x)$ diverges, which is why the latter is discarded when composing full-sky spherical harmonics.

Deviation from integer values of the degree $l$ and order $m$ amounts to re-evaluation of the boundary conditions of the Helmholtz equation solution at new locations, dictated by an observing mask that is separable along the angular and radial directions.
This ensures the SFB solutions are well-behaved and maintain orthogonality over the newly defined interval.
In practice while we allow for non-integer $l$ degree in this work, we keep the order $m$ integer-valued to simplify some of our calculations, as the Gaussian hypergeometric function has a useful simplification in this limit \citep{Cohl2020}.

As noted, the spherical cap mask is one example that has been widely used in the geophysics literature \citep{Haines1985}.
In this work, we introduce a SFB basis that is orthogonal over a 3D spherical stripe mask.
This mask is constructed by enacting a minimum and maximum $\theta$ and $\phi$ on the sphere, and a minimum and maximum $r$ along the line-of-sight.
In the formalism that follows, we leave the polar extent of the mask to be set arbitrarily by the observer (i.e. its polar arclength) but require that the azimuthal extent evenly divide into $2\pi$ (i.e. $2\pi/\Delta\phi \in \mathbb{N}$) for reasons discussed above.
In total, three parameters describe the observing mask: $\theta_{\rm min}$ is the minimum polar extent of the mask (i.e. the angle closest to the positive z axis); $\theta_{\rm max}$ is the maximum polar extent of the mask, and $\Delta\phi$ is the azimuthal extent of the mask, assuming that the mask starts at $\phi=0$.
Note this happens to be the observing mask of a fixed-pointing, drift-scan survey with a telescope at some designated latitude on Earth.

\begin{figure*}
\centering
\includegraphics[width=\linewidth]{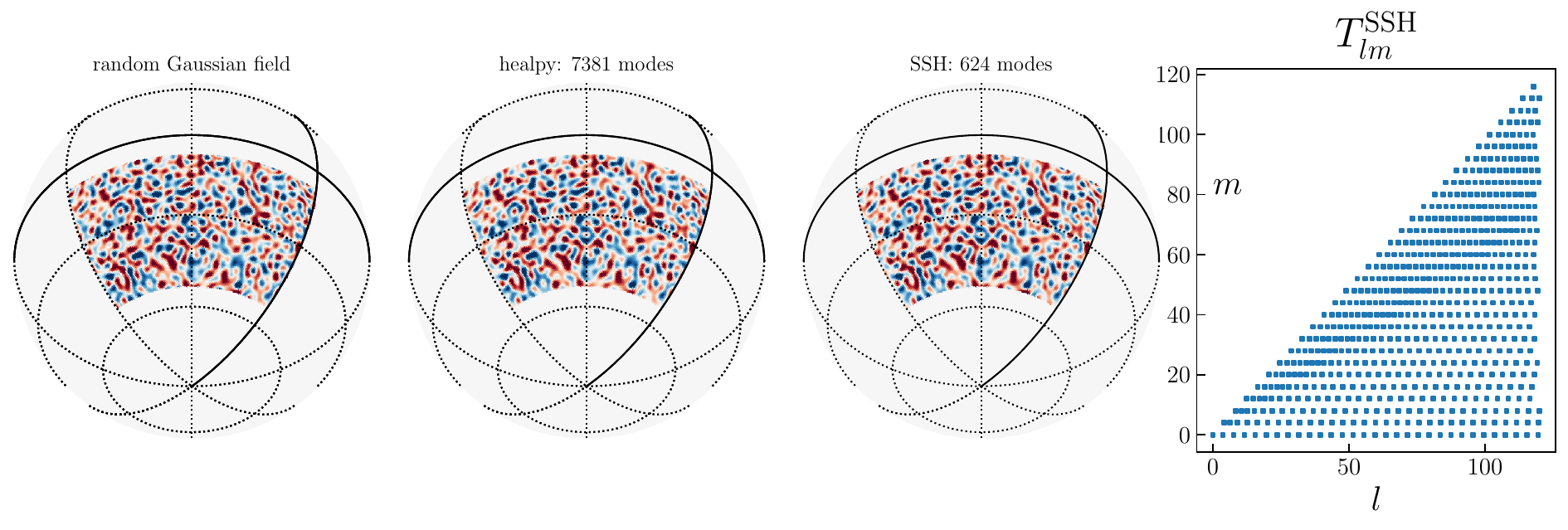}
\caption{Spherical harmonic fitting with full-sky spherical harmonics (healpy) and spherical stripe harmonics (SSH). We show a realization of a random Gaussian field on the cut sky (left), which is then transformed to the $a_{lm}$ domain using the full-sky routines and the cut-sky SSH routines with an $l_{\rm max}=120$, and then transformed back into map-space. The full-sky routines use over 7,300 modes for the given band-limit, while the SSH uses an order of magnitude fewer modes. We show the sampling points in $lm$ space of the SSH modes (right), highlighting their sparsity and non-uniform sampling of the $\ell$ and $m$ harmonic space.}
\label{fig:ssh_healpy}
\end{figure*}

Next, we derive the spectrum of $l, m,$ and $k$ modes than satisfy our boundary conditions along the polar, azimuthal, and radial axes, as well as maintain orthogonality over the spherical stripe mask.
These modes have the same behavior as full-sky spherical harmonics (i.e. composed of zonal, sectoral, and tesseral modes) and satisfy a set of boundary conditions set on the cut sky, however, they require a non-integer degree $l$ to do so.
The specifics of the stripe mask are described in \autoref{sec:ssfb_pspec}.
We also compare the new stripe decomposition against the standard full-sky routines in the \texttt{HEALPix} and \texttt{healpy} packages \citep{Gorski2005, Zonca2019}, which are widely used and well-tested.
We show the results of passing a simulated isotropic random Gaussian field cut with a spherical stripe mask on the sphere through the reverse and forward SHT of the full-sky (healpy) and the stripe harmonics (SSH) in \autoref{fig:ssh_healpy}, both with an $L_{\rm max}\sim120$. We show that the reconstruction of the field with the SSH is in good agreement with the full-sky routines, while using vastly fewer modes than their full-sky counterparts.
We also show the sampling of the $lm$ plane of the SSHs, demonstrating the sparse and non-uniform sampling that arises from conforming to the stripe mask.
For this mask, we acheive over an order of magnitude reduction in the number of parameters needed to model the signal up to the same bandlimit. 

\subsection{Azimuthal Boundary Conditions}
The azimuthal component of the general solution is written up to a constant as $\Phi_m(\phi) \propto e^{im\phi} + \tilde{A}^-_me^{-im\phi}$.
Given that the cosmological field is real-valued, negative $m$ modes contain no extra information so we can set $\tilde{A}_m^-=0$.
The boundary condition on $\Phi(\phi)$ is only that the function is periodic about the azimuth interval.
This implies that $e^{im0} = e^{im\Delta\phi}$, or that $m$ take on values $m=0/\Delta\phi,2\pi/\Delta\phi,4\pi/\Delta\phi$ and so on.  Given our condition on $\Delta\phi$, we see that $m \in \mathbb{N}$ as before, but now $m$ may not be separated simply by $\pm1$. For example, if a mask's azimuthal extent was $\Delta\phi=\pi/2$, then we get $m=0,4,8$ and so on.
Here we see that a truncation of the azimuthal range limits the number of $m$ modes generated for a given degree $l$: this is entirely expected, and in fact desireable, as it achieves one of the goals of this endeavor, which is to produce a sparser basis for modeling spherical signals on the cut-sky.
Note that an azimuthally truncated sky signal will not be inherehtly periodic over the interval $\Delta\phi$.
Although this is not ideal, it is not fundamentally different than the assumptions made in a discretized Cartesian Fourier basis.
The effect of an azimuthally non-periodic sky signal can be reduced by applying a tapering function that smoothly connects either end of the azimuthal range.
Finally, note that enforcing continuity over the interval $0<\phi<\Delta\phi$ imposes orthogonality between $\Phi_m(\phi)$ modes of different $m$, as is the case for the standard 1D Fourier basis.


\subsection{Polar Boundary Conditions}
The polar component of the general solution is written up to a constant as $\Theta_{lm}(\theta) \propto P_{l}^m(\cos\theta) + \tilde{A}^Q_{lm} Q_l^m(\cos\theta)$.
The standard conditions enforcing regularity for full-sky spherical harmonics are
\begin{align}
\frac{d\Theta_{lm}(\theta^\star)}{d\theta} = 0\hspace{.5cm} {\rm for}\ m &= 0\\
\Theta_{lm}(\theta^\star) = 0\hspace{.5cm} {\rm for}\ m &\neq 0,
\end{align}
where $\theta^\star$ is evaluated at the bounds of the interval at $\theta_{\rm min}=0$ and $\theta_{\rm max}=\pi$.
For \emph{spherical cap harmonic} formalism \citep{Haines1985, Thebault2006, Torta2019}, we shift the maximum polar boundary to some arbitrary value within $0<\theta_{\rm max}<\pi$ and set a boundary condition on either the field or its derivative at the surface edge, also known as Dirichlet (Type 1) or Neumann (Type 2) conditions, respectively.
This condition can be written as
\begin{align}
\label{eq:sca_bcs}
C_1\Theta_{lm}(\theta_{\rm max}) + C_2\Theta_{lm}^\prime(\theta_{\rm max}) = 0,
\end{align}
where one would set $C_2=0$ for Dirichlet conditions or $C_1=0$ for Neumann conditions, and $\Theta_{lm}^\prime$ is the polar derivative of $\Theta_{lm}$.
The benefit of Neumann conditions is the ability to reconstruct an arbitrary signal amplitude when approaching the surface edge.
Because $Q_{l}^{m}$ diverges at $\theta=0$ it is not included in the standard spherical harmonics or the spherical cap harmonics.

For the \emph{spherical stripe harmonics} presented in this work, the boundary condition at $\theta_{\rm min}$ takes the same form as $\theta_{\rm max}$.
However, in order to satisfy both boundary conditions we must now include $Q_l^m$ terms into the solution.
Assuming, for simplicity but without loss of generality, Neumann conditions of $C_1=0$, the spherical stripe boundary conditions lead to the equality
\begin{align}
\label{eq:polar_stripe_equalities}
P_l^{\prime m}(\cos\theta^\star) + \tilde{A}_{lm}^{Q}Q_l^{\prime m}(\cos\theta^\star)=0,
\end{align}
where $\theta^\star$ is evaluated at $\theta_{\rm min}$ and $\theta_{\rm max}$, and $P_l^{\prime m}$ is the polar derivative of $P_l^m$.
This can be rearranged to form
\begin{align}
\tilde{A}_{lm}^{Q} = \frac{-P_l^{\prime m}(\cos\theta^\star)}{Q_l^{\prime m}(\cos\theta^\star)},
\end{align}
where $\cos\theta^\star$ can be either $\cos\theta_{\rm min}=x_{\rm min}$ or $\cos\theta_{\rm max}=x_{\rm max}$,
and
\begin{align}
&P_l^{\prime m}(x_{\rm min})Q_l^{\prime m}(x_{\rm max}) - P_l^{\prime m}(x_{\rm max})Q_l^{\prime m}(x_{\rm min}) = 0.
\end{align}
The latter is used to solve for the non-integer spectrum of $l$ for each integer $m$, and the former is used to solve for the $\tilde{A}_{lm}^Q$ coefficients for each $l$ and $m$.
Normalization of the spherical harmonics $Y_{lm}(\theta,\phi)=\Theta_{lm}(\theta)\Phi_m(\phi)$ can then be performed such that their inner product sums to one.

Computing the associated Legendre functions $P_l^m(x)$ and $Q_l^m(x)$ on the interval $x\in[-1, 1]$ for high orders ($l=m>100$) with standard formulas for the Gaussian hypergeometric function \citep{Cohl2020} will result in numerical overflow even with double precision arithmetic. This is a result of the large dynamic range in the normalization factors that scale as $m!$ and $(l+m)!$, even though they effectivly cancel out in the final spherical harmonic.
To alleviate this, one simply needs to scale down the hypergeometric function by $1/|m|!$ when computing it, and then re-normalize when applying the overall orthonormalization factor. Furthermore, evaluating factorial or gamma function products with large arguments in log space, summing and subtracting neighboring log factorials, and then exponentiating will help to prevent numerical overflow.
In \texttt{BayesLIM}, these implementations allow for stable computation of high integer order and non-integer degree spherical harmonics up to $l=m\sim400$, which is sufficiently high for the studies performed in this work.


\subsection{Radial Boundary conditions}

The general solution for the radial component is written up to a constant as $g_l(kr) \propto j_l(kr) + \tilde{A}^y_{l}y_l(kr)$.
There are multiple methods for enacting boundary conditions along the radial dimension of the SFB, including imposing the field or its derivative go to zero at the boundary \citep[e.g.][]{Heavens1995, Leistedt2011, Samushia2019, Chakraborty2019}, or imposing continuity on a potential field and its gradient \citep[e.g.][]{Fisher1995, Gebhardt2021}.
Doing so discretizes the $k$ wavevectors into a spectrum of $k_{n}$ orthogonal modes along the radial interval.
Additionally, similar to the polar coordinate, we can choose to truncate the radial mask and enact the boundary conditions at an arbitrary $r_{\rm min}$ and $r_{\rm max}$.
While many previous works have assumed the case of $r_{\rm min}\to0$ for simplicity, recent studies have begun to derive boundary conditions for a non-trivial $r_{\rm min}$ \citep{Samushia2019, Chakraborty2019, Gebhardt2021}.

For an intensity mapping survey with a well-defined radial selection function (i.e. the set of observed frequencies), it is fairly straightforward to take the latter approach.
Imposing (Neumann) conditions at the radial boundaries yields
\begin{align}
j_l^\prime(k_{ln}r^\star) + \tilde{A}_{ln}^y y_l^\prime(k_{ln}r^\star) = 0,
\end{align}
where $r^\star$ is either $r_{\rm min}$ or $r_{\rm max}$.
Similar to the polar axis case, we see that both the first and second order spherical Bessel functions are needed when using a non-vanishing $r_{\rm min}$.
This yields the following equalities,
\begin{align}
\tilde{A}_{ln}^y = \frac{-j_l^\prime(k_{ln}r^\star)}{y_l^\prime(k_{ln}r^\star)},
\end{align}
and
\begin{align}
j_l^\prime(k_{ln}r_{\rm min})y_l^\prime(k_{ln}r_{\rm max}) - j_l^\prime(k_{ln}r_{\rm max})y_l^\prime(k_{ln}r_{\rm min}) = 0.
\end{align}
The zeros of the latter equation yields the spectrum of $k_{ln}$ modes where our radial boundary conditions are satisfied, and the former can be used to compute the relative coefficient of the spherical Bessel functions. Finally, we can re-normalize the radial modes such that their inner products satisfy our previous orthogonality condition (\autoref{eq:SBT_ortho}).

\subsection{Power Spectrum Estimation}
\label{sec:ssfb_pspec}

The 3D spatial Fourier transform of the temperature field is defined as
\begin{align}
\label{eq:3d_FT}
\widetilde{T}(\bk) = \int d^3r\ e^{i\bk\br}T(\br)
\end{align}
with its inverse transform defined as
\begin{align}
\label{eq:3d_iFT}
T(\br) = \int \frac{d^3k}{(2\pi)^3}\ e^{-i\bk\br}\widetilde{T}(\bk).
\end{align}
The relationship between the field in Fourier space and its power spectrum is
\begin{align}
\langle \widetilde{T}(\bk)\widetilde{T}^\ast(\bk^\prime)\rangle = (2\pi)^3 \delta^D(\bk-\bk^\prime)P(\bk),
\end{align}
where $\langle\rangle$ represents an ensemble average and $\delta^D(\bk-\bk^\prime)$ is the Dirac delta function.
To estimate the power spectrum with the SFB or SSFB formalism, we need a relation between $P(k)$ and the SFB coefficients of \autoref{eq:sfb_reverse}, which is given in \citep{Liu2016c} as
\begin{align}
\label{eq:sfb_pspec}
\langle T_{lm}(k)T_{l^\prime m^\prime}^\ast(k^\prime)\rangle = k^{-2}\delta^D(k-k^\prime)\delta_{ll^\prime}\delta_{mm^\prime}P(k).
\end{align}
This demonstrates the close relationship between $\widetilde{T}(k)$ and $T_{lm}(k)$ modes and their connection to the power spectrum.
This also shows us that the averaged, 1D power spectrum can be recovered by binning $T_{lm}(k)$ in $k$ and averaging over all $l$ and $m$ modes; however, \citep{Liu2016c} points out that for a practical survey, the minimum variance estimate of the power spectrum requires an $l$ (and possibly $m$) dependent weight in the average to account for the survey geometry.

To demonstrate the SSFB formalism in practice, we apply it here to a simulated wide-field 21\,cm data cube using the quadratic estimator (QE) formalism \citep{Hamilton1997, Tegmark1997, Liu2014a, Dillon2015a}.
For clarity, we will first briefly describe the quadratic estimator formalism: a full review is beyond the scope of this section and curious readers should consult the above references.

To make a numerical estimate of an underlying continuous power spectrum, we can discretize the power spectrum into a set of \emph{band powers}.
To do so, we model the continuous power spectrum as piecewise constant whose amplitudes in each $k$ band are called the band powers,
\begin{align}
P(k) = p_1\Gamma_1(k) + p_2\Gamma_2(k) + \ldots = \sum_\alpha p_\alpha \Gamma_\alpha(k),
\end{align}
where $\Gamma_\alpha(k)$ is the windowing function of the $\alpha$ band power, which is 1 for all $k$ within $k_\alpha^{\rm min}<k<k_\alpha^{\rm max}$ and 0 otherwise.
Notably, the band powers are linearly related to the covariance matrix of the data as
\begin{align}
\label{eq:bandpower}
C(\br,\br^\prime) = \langle T(\br)T(\br^\prime)^\ast\rangle = \sum_\alpha p_\alpha C_{,\alpha}(\br,\br^\prime),
\end{align}
where $C_{,\alpha}$ is the derivative of the covariance with respect to the $\alpha$ band power.
If we discretize the temperature field (i.e. the data) into a column vector $\bx$ of length $NM$ where $N$ is the number of radial shells and $M$ is the number of sky pixels, and discretize the SFB transform into a row vector $\bz$ of the same size,\footnote{In discretizing the SFB transform, we are choosing to sample the transform integrand at the centroid of each pixel, and then approximate the integral as a discrete sum. In principle we can account for the effects of the pixelization by adopting a pixel window function \citep[e.g.][]{Dillon2014}, but for now we deem that beyond the scope of this simple demonstration.} we can express \autoref{eq:sfb_reverse} as
\begin{align}
t_{lmn} = \bz_{lmn} \bR \bx,
\end{align}
where $\bz_{lmn}$ is a row vector, $t_{lmn}$ is a single $T_{lm}(k_n)$ coefficient, and $\bR$ is a square matrix that acts as a pre-weighting of the data before taking the SFB transform. This can, for example, be a diagonal matrix that holds an apodization or windowing of the data such that the data transition smoothly to the mask boundaries.
Throughout this section, we use lowercase boldface to denote vectors and uppercase boldface to denote matrices.

Drawing from our conclusion that the power spectrum is related to the square of $T_{lm}(k)$, we assert that an (unnormalized) estimate of the power spectrum can be written as
\begin{align}
\label{eq:unnorm_q}
\hat{q}_{lmn} = t_{lmn}^\ast t_{lmn}=\bx^\dagger \bR^\dagger \bz_{lmn}^\dagger \bz_{lmn} \bR \bx,
\end{align}
where $\hat{q}$ implies it is an estimate of $q$ from our finite survey volume.
One may notice the similarity of \autoref{eq:unnorm_q} to that of the quadratic estimator \citep{Tegmark1997, Liu2011}, where $\bz_{lmn}^\dagger\bz_{lmn}$ is equivalent to $C_{,\alpha}$ discussed before and $\alpha$ indexes a unique $lmn$ combination.
We will use this similarity to derive certain statistical properties of the estimated power spectrum, specifically its window functions.

\begin{figure*}
\centering
\includegraphics[width=\linewidth]{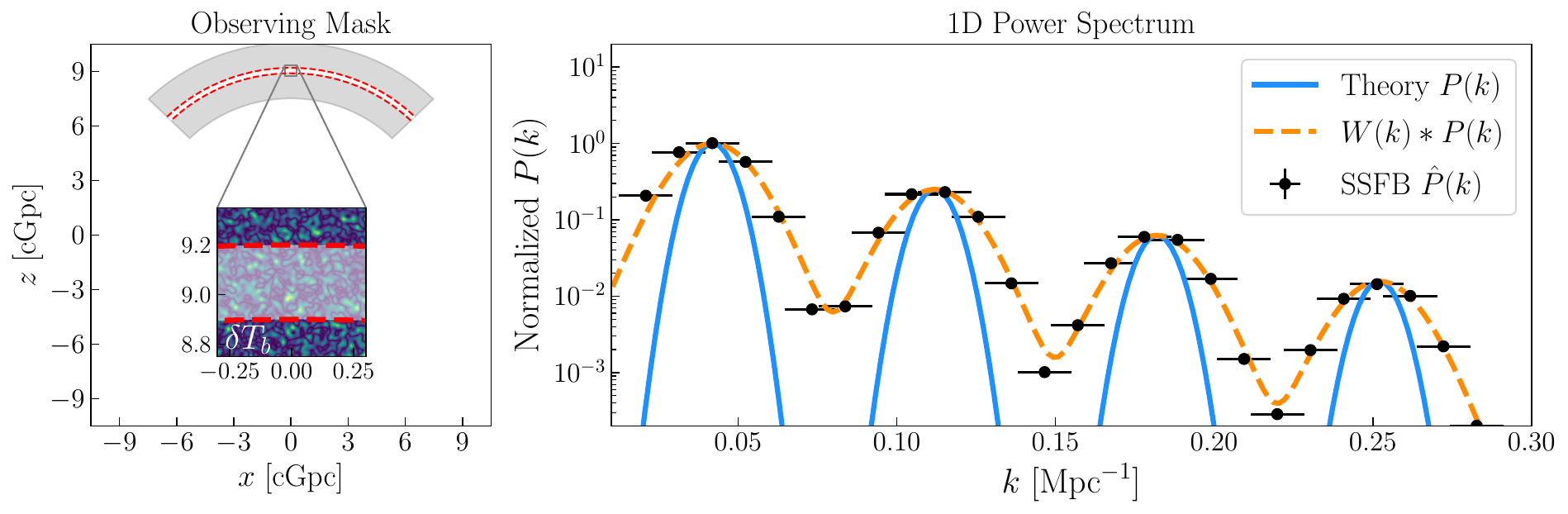}
\caption{SSFB power spectrum recovery test of a simulated random Gaussian field. We show the azimuthal and radial extent of the observing mask (left), where $z$ is oriented along the line-of-sight, with a zoom-in inset showing the simulated random Gaussian field. We also show the recovered SSFB power spectrum (right) with their errorbars (black points), the theoretical input, double-tone power spectrum (blue), and the theoretically measured power spectrum (dashed orange), which is the theory $P(k)$ convolved with the SSFB estimator's window function. The measured points are in good agreement with the convolved theory prediction, with small vertical errorbars given the large angular extent of the mask. We see accurate recovery of the location of the tones in $k$ space, as well as the relative amplitudes of the tones. All curves have been normalized by the peak of the theory curve (blue) to better capture the dynamic range as a function of $k$.}
\label{fig:ssfb_pspec}
\end{figure*}

Following the quadratic estimator formalism, we introduce a normalization matrix $\bM$ to produce a normalized estimate of the power spectrum,
\begin{align}
\hat{p}_{\alpha} = \sum_\beta M_{\alpha\beta} \hat{q}_{\beta}.
\end{align}
Note that this does not show the usual bias term associated with the QE because we are assuming, for the sake of this demonstration, that the data is only populated by a signal term (no noise or foregrounds).
Following \citet{Dillon2014}, one can also ignore the bias terms if we use statistically independent samples for $\bm{x}$ in \autoref{eq:unnorm_q}, which is often the case in real analyses \citep{HERA2022a}.
In case one isn't using statistically independent noise draws, there will be a noise bias term, which can be estimated from the input noise covariance and subtracted.

Taking the expectation value of our estimated power spectrum yields
\begin{align}
\langle \hat{p}_{\alpha}\rangle &= \sum_{\alpha\beta}M_{\alpha\beta}\langle \hat{q}_{\beta}\rangle \nonumber\\
&= \sum_\beta M_{\alpha\beta} \Tr[\langle\bx\bx^\dagger\rangle\bR^\dagger\bz^\dagger_\beta\bz_\beta\bR] \nonumber\\
&= \sum_\beta M_{\alpha\beta}\sum_{\gamma}\Tr[\bz^\dagger_\gamma\bz_\gamma\bR^\dagger\bz^\dagger_\beta\bz_\beta\bR]p_\gamma \nonumber \\
&= \sum_{\beta\gamma}M_{\alpha\beta}|\bz_\beta\bR\bz_\gamma^\dagger|^2p_\gamma \nonumber \\
&= \sum_{\beta\gamma}M_{\alpha\beta}H_{\beta\gamma}p_\gamma,
\end{align}
where in the third line we used \autoref{eq:bandpower}, and in the fourth line we recognized that $\bz_\beta\bR\bz_\gamma^\dagger$ is a scalar and $\bR=\bR^\dagger$.
This final form reveals that the estimated bandpowers are related to the true bandpowers via a window function matrix $\bW$ defined as
\begin{align}
\langle\hat{\bp}\rangle = \bM\bH\bp = \bW\bp.
\end{align}
Various choices of the normalization matrix $\bM$ can be made that yield different properties of the estimated bandpowers \citep[e.g.][]{Tegmark2002}: the only constraint is that we choose an $\bM$ matrix that allows the rows of $\bW$ to sum to one, thus making $\hat{\bp}$ an unbiased estimator \citep[see also][]{Liu2011, Liu2014a, Dillon2015a, Kern2021}.
In this work, we will adopt a diagonal $\bM$ matrix that simply enforces this property for each row in $\bW$.

Note that because of the orthogonality of $f_{lmn}$ with respect to $l^\prime$, $m^\prime$, and $n^\prime$ over the observing mask, one can show that the off-diagonal of $H_{\alpha\beta}$ vanish while the diagonal holds the inner product of $f_{lmn}$, making the window function also diagonal and therefore trivial. 
However, orthogonality is partially broken in the case a non-uniform pre-weighting $\bR$ matrix.
In this work, we apply a Hann tapering function across the line-of-sight axis, meaning the only non-trivial components of the window function are those of $\bW_{n}^{n^\prime}$, which are considerably smaller and easier to compute than the general $\bW_{lmn}^{l^\prime m^\prime n^\prime}$.
This tapering (or apodization) is applied to reduce sidelobe $k$-mode ringing due to the fact that the real data do not strictly meet the boundary conditions assumed by the radial modes (either that the field or its derivative goes exactly to zero).

Next we demonstrate the ability of the SSFB approach to accurately model and re-construct a random Gaussian field observed through a spherical stripe observing mask, and ensure that it produces an accurate estimate of the field's power spectrum.
The mask used in this test extends across the azimuthal direction $-45^\circ<\phi<45^\circ$, the polar direction $35^\circ<\theta<80^\circ$, and the radial direction $8900<r<9200$ cMpc, which corresponds to a frequency range of 155--173 MHz and a redshift range of 7.2--8.2 for the 21\,cm line.
We simulate a 3D periodic box with $500^3$ voxels and a sidelength of $L=1000$ cMpc (2.0 cMpc resolution).
The random field is drawn from a power spectrum with a decaying set of tones, with a maximum $k$ scale of 0.3 cMpc$^{-1}$.
This test seeks to ensure that 1) the $k$ modes of the tones are recovered accurately and 2) the relative amplitude between the tones are recovered.

After generating a 3D Gaussian random field we tile it onto an NSIDE=512 HEALpix map at 64 different shells within the radial range.
We do this by stacking the boxes in 3D out to the spherical shell comoving radius and use bilinear interpolation to sample each map pixel.
We then band-limit the maps by smoothing them at an $l_{\max}=90$, and then downsample them to a NSIDE=128 HEALpix resolution.
Note this is similar to how the 21\,cm signal simulation is constructed in \autoref{sec:HI_signal}.

The result of estimating the SSFB power spectrum on the simulations described above is shown in \autoref{fig:ssfb_pspec}, which shows a cut through the survey mask (left) where $z$ is oriented along the line-of-sight and the zoom-in inset shows the a slice of the simulated random Gaussian field. The gray shaded region shows the full extent of the volume probed by an experiment like HERA (50 - 250 MHz), while the red dashed box shows the region over which the power spectrum is actually estimated. The estimated spherical stripe Fourier Bessel (SSFB) power spectrum is plotted (right, dots) against the input quadruple-tone theory power spectrum (blue) and the theory power spectrum convolved with the SSFB estimator's window function (dashed orange).
Horizontal errorbars represent the full width half max of the window functions, vertical errorbars (not visible) represent sample variance on the signal given the finite volume.
Relative to the convolved theory (dashed-orange), the SSFB estimator does a good job reconstructing the power spectrum, and accurately measures the location and relative amplitude of the inserted tones in the power spectrum.
We artifically normalize the unitless power spectra by the peak theory curve (blue) to better capture the dynamic range as a function of $k$.

\section{Efficient Matrix-Vector Products for Dense Mass Matrices in HMC}
\label{sec:hmc}

Recall that the Hamiltonian Monte Carlo (HMC) approach to posterior sampling uses Hamiltonian dynamics to trace the trajectory of a particle in a potential well defined by the negative log posterior distribution \citep{Duane1987, Neal2011, Betancourt2017}.
To briefly review our notation, we define the particle position and momentum column vectors as $\bq$ and $\bp$, respectively, where the position vector is a proxy for the forward model's parameter vector.
The covariance of the position vector is $\bC$ and its inverse is called the \emph{mass matrix} $\bM$, which is equivalent to the model's Hessian matrix (the matrix containing the posterior's second derivatives with respect to the model parameters).
We will define a lower-triangular Cholesky decomposition of the mass matrix as $\bM = \bL_M \bL_M^T$.
The key quantities that are required to accurately simulate this Hamiltonian trajectory are a series of matrix-vector products, including:
\begin{itemize}
\item[1.] $K = \tfrac{1}{2}\bp^T\bM^{-1}\bp$ [kinetic energy term]
\item[2.] $\partial_t \bq = \bM^{-1}\bp$ [position update term]
\item[3.] $\bp = \bL_M \bp_0$ [momentum scaling term],
\end{itemize}
which can be found in Eqn. 2.6, Eqn. 2.7, and Sec. 4.1, respectively, from \citet{Neal2011}.
In many cases, the mass matrix is approximated as diagonal, which simplifies the above matrix-vector products into trivial element-wise vector operations. However, for poorly conditioned posteriors this can make sampling extremely inefficient, and a dense mass matrix can dramatically improve sampling efficiency if such matrix operations can be computationally tolerated.
However, even if we can compute and store the mass matrix, we generally will not want to invert it, as would be suggested by the above equations.
General matrix inversion scales as $\mathcal{O}(N^3)$ and can be unstable depending on the matrix's condition number.
Instead, we can use relationships between the previously defined Cholesky factors and use efficient triangular linear solves, which run in $\mathcal{O}(N^2)$ time.
Note that although Cholesky factorization also scales as $\mathcal{O}(N^3)$ it has a smaller prefactor (roughly 3$\times$ faster than inversion), and is more stable than direct inversion \citep{Nocedal}.

One caveat is that the Hessian matrix may not necessarily be positive definite (e.g. if we are at a saddle point), which would prohibit a Cholesky factorization.
To work around this, we can use the nearest positive definite approximation of the Hessian by regularizing it, which fits well with our Bayesian approach because this is equivalent to placing a stronger prior on our parameters.
The total Hessian of the negative log posterior is simply the Hessian of the negative log likelihood summed with the Hessian of the negative log prior.
To make the minimal adjustment needed to make the posterior Hessian symmetric positive definite (SPD), we experiment by adding small multiplicative increases to the computed prior Hessian until the posterior Hessian becomes SPD.

Next, given a permissible Cholesky factorization of the mass matrix, we will briefly show how we can compute the three required quantities for simulating HMC.
First, we relate the inverse of the mass matrix to it's Cholesky factors
\begin{align}
\bM^{-1} = \bL_M^{-T}\bL_M^{-1}.
\end{align}
Then, we can rewrite the kinetic energy term (1.) as
\begin{align}
K = \tfrac{1}{2}(\bL_M^{-1}\bp)^T \bL_M^{-1}\bp = \tfrac{1}{2}\bz^T\bz.
\end{align}
This means we can efficiently compute $\bz$ via forward substitution of the linear system,
\begin{align}
\bL_M \bz = \bp.
\end{align}

Next, we can define a similar solution for the position update term (2.), which uses forward substitution followed by backward substitution.
We can rewrite (2.) above as
\begin{align}
\partial_t\bq=\bM^{-1}\bp = \bL_M^{-T}\bL_M^{-1}\bp = \bL_M^{-T}\bz.
\end{align}
We can first use forward substitution to solve $\bL_M\bz=\bp$ for $\bz$, then we can use backward substitution to solve $\bL_M^T(\partial_t\bq)=\bz$.
Finally, computing (3.) is straightforward, where $\bp_0\sim\mathcal{N}(0,1)$.
A Hessian-preconditioned HMC sampler was also proposed in \citet{Lentati2013, Sims2019}; however, their approach uses an eigendecomposition of $\bm{M}$, which has a larger upfront cost than Cholesky factorization but also has the advantage of being able to work directly with singular mass matrices.

Relatedly, we can also draw uncorrelated samples from the parameter covariance $\bC$ given only access to $\bL_M$.
To draw a random sample $\bv\sim\mathcal{N}(0,\bC)$, usually we would first draw an uncorrelated unit-Gaussian vector $\bv_0\sim\mathcal{N}(0,1)$ and then transform it by the Cholesky of the covariance.
However, using the relationship above, we can also solve for $\bv$ via backward substitution of
\begin{align}
\bL_M^T \bv = \bv_0,
\end{align}
where $\bL_M^T$ is upper triangular.
Note that these are not draws from the true posterior, but are draws from a covariance defined implicitly by the mass matrix, which is an approximation to the true posterior (also known as the Laplace approximation).

We show the MCMC chains from our proof-of-concept run for a handful of parameters along with their averaged autocorrelations in \autoref{fig:hmc_autocorr}.
Note that, as discussed in \autoref{sec:hmc}, the HMC sampler is preconditioned with a block diagonal mass matrix, where each component in our data model (EoR, foreground, beam) is assumed to be dense, but the inter-component off-diagonals are zero (with the exception of the foreground-beam off-diagonals, which are kept for reasons discussed in \autoref{sec:sampling}).

To assess how many independent samples we have drawn from the posterior we can compute the effective sample size (ESS).
Because MCMC chains inevitably have some amount of correlation between samples, the effective sample size is generally less than the total sample size.
The ESS is therefore defined as the chain length ($N$) divided by the average autocorrelation of the chain, computed as
\begin{align}
\label{eq:ess}
N_{\rm eff} = \frac{N}{1 + 2 \sum_{\tau=1}^M\rho(\tau)},
\end{align}
where $\rho(\tau)$ is the measured autocorrelation function of the chain as a function of the sample lag ($\tau$), and the sum runs up to an integer $M < N$ to limit sampling noise in $\rho(\tau)$ from affecting our $N_{\rm eff}$ estimate \citep{Vehtari2021}.
For our MCMC chains shown in \autoref{fig:hmc_autocorr}, we compute the effective sample size by first taking the average of all measured autocorrelations functions within each component, and then use \autoref{eq:ess} to derive $N_{\rm eff}$ for the averaged autocorrelation function (dashed lines in \autoref{fig:hmc_autocorr}).
We compute effective sample sizes of 3, 5, \& 15 for our EoR, foreground, \& beam components, respectively, out of our $N\sim500$ length chains.
We suspect the beam component has a longer autocorrelation length because the underlying parameterization is more internally degenerate and more poorly conditioned than that of the foreground and EoR components.

One twist we added to make the sampling more efficient is an adaptive step size feature.
Before sampling, we can tune the HMC step size to yield high acceptance probability, which we can do manually or automatically via a dual-averaging approach \citep{Hoffman2011}.
However, for complex distributions, for example ones that are not simply multivariate Gaussians, the sampler can walk into regions of parameter space with sufficiently higher curvature leading to larger HMC integration errors that force down the acceptance rate to very low levels.
HMC step size adaptation is a means for trying to automatically adjust the step size to account for regions of higher curvature, such as the delayed rejection approach \citep{Modi2024}.
Here, we use a similar but slightly different step size adjustment approach.
Let the originally-tuned step size parameter be $\epsilon_0$.
After simulating an HMC trajectory, if the acceptance probability of the final state falls below some pre-defined threshold (say 0.2) then we shrink the step size parameter by $\sim20\%$ of its current value: in other words, we set $\epsilon\leftarrow\epsilon/1.2$.
A new trajectory is then proposed and integrated and we repeat the update process, with a key difference being that the step size does not refresh to its original value, but keeps its reduced size.
At the same time, for future trajectories, if the acceptance probability falls above the pre-defined threshold, we increase the step size by $\sim20\%$ of its current value with a maximum achievable value of its original setting: in other words, $\epsilon\leftarrow{\rm Min}[\epsilon_0, 1.2\epsilon]$.
Thus our approach can be thought of as a running step size adjustment that tracks the sampler as it walks through the parameter space.
This is by no means an optimal step size adjustment procedure necessarily, but one that worked for the proof-of-concept at hand.
Future work will aim to incorporate more complex and dynamic step size and mass matrix adaptations as needed.

\section{Computational Scaling}
\label{sec:scaling}

\begin{figure*}
\includegraphics[width=.32\linewidth]{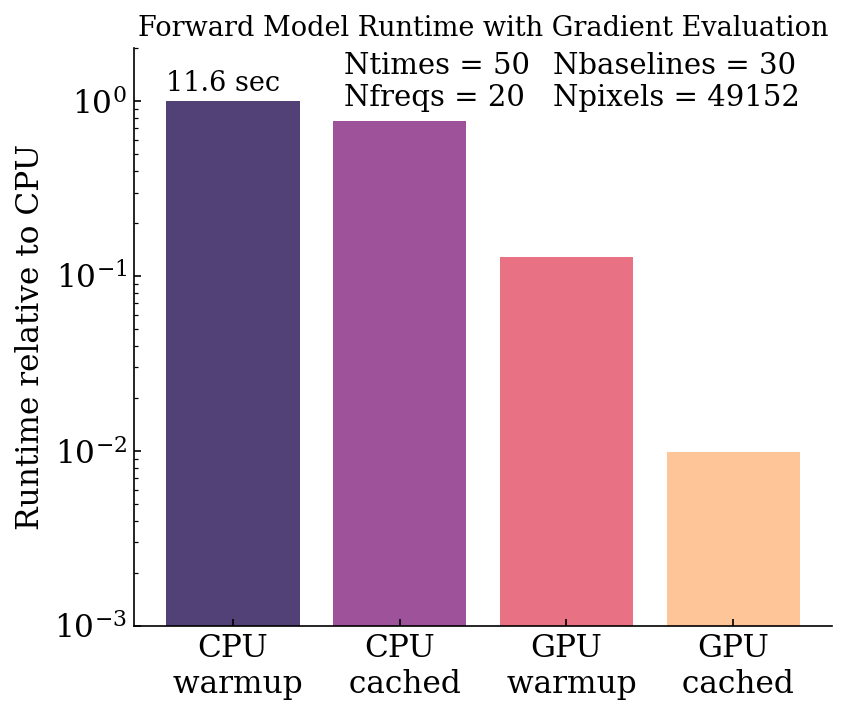}
\includegraphics[width=.32\linewidth]{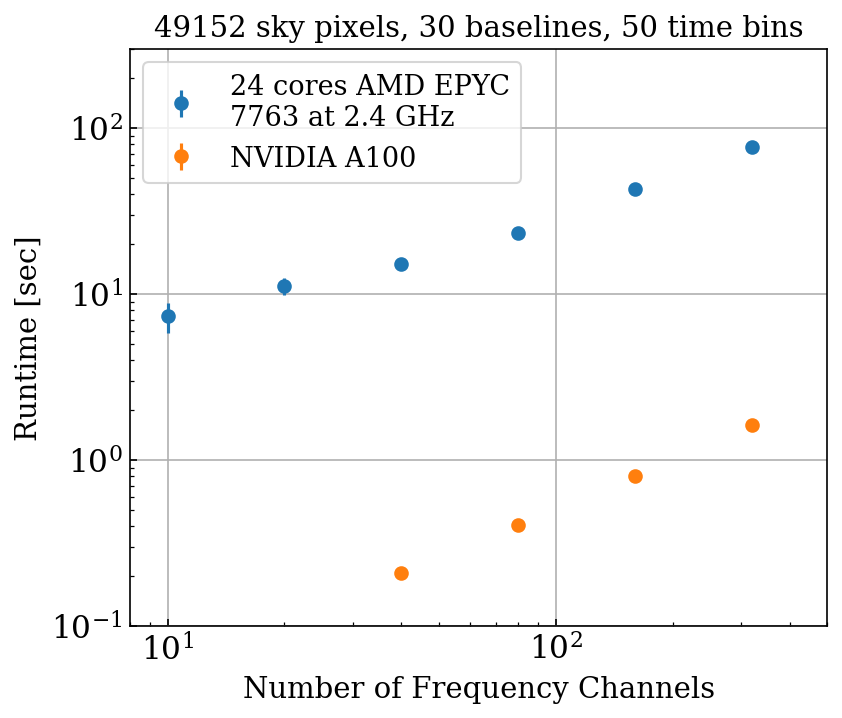}
\includegraphics[width=.32\linewidth]{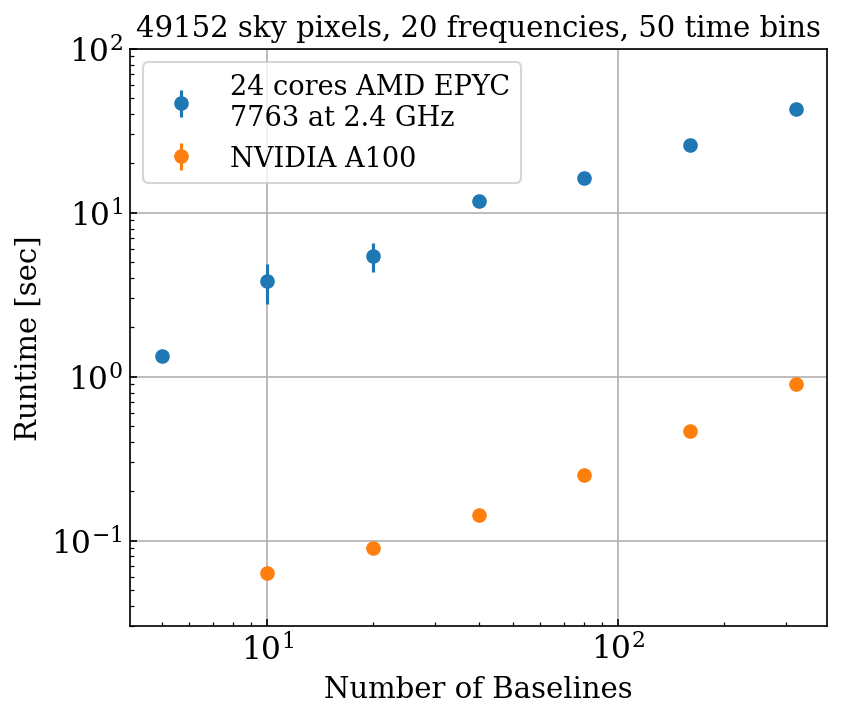}
\caption{Computational benchmark and scaling tests of the \texttt{BayesLIM} forward model. The model used in these tests include an NSIDE 64 resolution sky map and a 1-degree resolution antenna primary beam model (hemispherical), where the parameters for each component are the pixel values.
We simulate 20 frequency channels, 50 time bins, and 30 baseline vectors, and profile the total runtime of the foward pass and the backpropagation step.
We run the profiling on a 24-core AMD EPYC 7763 CPU running at 2.4 GHz as well as a single NVIDIA A100 GPU.
We find that roughly equal time is split between the forward pass and the backpropagation step.
Left: The runtime of the model (forward pass and backpropagation step) on the CPU and GPU in a warm-up and a cached mode.
Relative to the CPU timing, we see that the GPU in the cached mode delivers a factor of 100 in speed-up.
Center: A scaling test showing the runtime in cached mode while varying the number of frequency channels.
The plot the average and standard deviation of 20 runs for each scenario.
We see the the runtime approach linear scaling.
Right: A similar scaling test showing the runtime relative to the number of baselines in the data. Again we see similar speed-ups for the GPU relative to the CPU and linear  scaling with an increased number of baselines.
}
\label{fig:benchmarks}
\end{figure*}

Given the computational demands of the proposed framework, we run benchmarks and scaling tests to forecast the required computational load for an analysis on real data where the number of time stamps, frequencies, or baselines could be non-negligibly larger.
In \autoref{fig:benchmarks} we show the results of CPU and GPU based benchmarks, as well as scaling tests while varying the number of frequency channels and baselines being computed.
For the scaling tests, we plot the mean and standard deviation of 20 runs for each test case.
The model adopted for these tests include a single NSIDE 64 resolution sky model ($\sim10^4$ sky pixels), whose parameters are each sky pixel for each frequency channel, and a 1-degree resolution antenna primary beam model, whose parameters are each sky pixel for each frequency channel.
We use a default of 30 baselines, 50 time bins, and 20 frequency channels, unless otherwise specified.
The array model adopted is a HERA-217 array, but this is not actually relevant to the runtime of the tests, as what really matters is the number of unique baselines being simulated $(N_{\rm baselines}$), which is explicitly controlled for in the tests.
Therefore, if we specify $N_{\rm baselines}=30$, this means we only simulate 30 baselines of the total number of unique baselines in the array, and the choice of baseline has no impact on the runtime.
We run the profiling on 24 cores of a single AMD EPYC 7763 CPU clocked at 2.4 GHz, in addition to a single NVIDIA A100 GPU with 80 GB of VRAM.
In both cases we run the model in double precision.

We profile the runtime of a single gradient update, which involves the visibility simulation forward pass and the computation of the parameter gradients via backpropagation.
For the model adopted here, roughly equal time is spent in the forward pass and the backward pass.
For our first test (Left, \autoref{fig:benchmarks}), we show the runtime on the CPU and the GPU in two modes.
The first is a ``warmup'' mode where we have yet to compute intermediate products necessary for the forward pass (e.g. coordinate transformations, beam interpolation splines, etc.).
These, it turns out, are often the bottleneck for realistic RIME visibility simulations \cite{Kittiwisit2025}.
The second mode, called a ``cached'' mode, is profiled where we cache all of these intermediate products so that they can be automatically resused.
This allows the forward pass to be largely dominated by the matrix operations described by \autoref{eq:rime_matrix}, which allows the GPU to deliver significant acceleration.
Thus, we see nearly two orders of magnitude in speed-up delivered by the GPU relative to the CPU when operating in the cached mode (which is the normal operating mode after a single forward pass).

Next we show the results of model scaling with respect to the number of frequency channels and the number of simulated baselines (center and right, 
\autoref{fig:benchmarks}).
We plot the average and standard deviation of 20 runs for each test case.
We expect to see roughly linear scaling with the number frequencies and baselines \citep{EwallWice2022, Kittiwisit2025}, which we observe in both cases.
For these tests we are always operating in the cached mode, and again observe a speed-up on the GPU by over an order of magnitude relative to the CPU.
Assuming continued linear scaling, these benchmarks put \texttt{BayesLIM} on-par in terms of speed with other state-of-the-art, GPU-based visibility simulators that are being developed for next-generation radio telescopes like the SKA \citep{Kittiwisit2025, Ohara2025}.
Extrapolating these benchmarks to a realistic data quantity for HERA Phase II \citep{Berkhout2024}, we estimate that a similar analysis as demonstrated here but with double the the number of frequency channels and time integrations could be run in under a few hundred GPU hours, which is a very reasonable cost given the availability of GPU compute.

One of the main limitations for this approach is the large amount of GPU VRAM needed.
For the proof-of-concept described here, the VRAM needed to hold the entire model and its graph in memory exceeds 200 GB.
However, there are many approaches for working around this limitation.
First, we can split the data across multiple GPUs in a data-parallel manner to lower the VRAM needed for each individual GPU.
Assuming we have a fast GPU-to-GPU interconnect, this allows for efficient parallelization across GPUs.
Second, we can further split the data into minibatches, run a forward and backward pass on each minibatch sequentially (deleting the computational graph after each minibatch), and sum the gradients across minibatches before taking a parameter update step, commonly known as gradient accumulation.
This lowers the peak instantaneous VRAM needed, allowing us to use lower-memory GPUs, at the cost of some redundant operations and thus a slower runtime.
We employ all of these techniques to make this proof-of-concept work on a set of 4xA100 GPUs running simultaneously on a single computational node.


\bsp	
\label{lastpage}
\end{document}